\documentclass{article}
\usepackage{amsmath}
\usepackage{lipsum} 
\usepackage[top=2.5cm, bottom=3cm]{geometry}
\usepackage{geometry}
\usepackage[utf8]{inputenc}
\usepackage{xcolor}
\usepackage{booktabs}
\usepackage{setspace}
\usepackage{graphicx}
\usepackage[hidelinks]{hyperref}
\usepackage[document]{ragged2e}
\usepackage{pythonhighlight}
\usepackage{pdflscape}
\usepackage{longtable}
\usepackage{epstopdf}
\usepackage[flushleft]{threeparttable}
\usepackage{comment}
\usepackage{tikz}
\usepackage{standalone}
\usepackage{array}
\usepackage{adjustbox}
\usepackage{svg}
\usepackage{dutchcal}
\usepackage{changepage}
\usepackage{dutchcal}
\usepackage{caption}
\usepackage{subcaption}
\usepackage{soul}
\usepackage{float}
\usepackage{mdframed}
\usetikzlibrary{mindmap,trees}
\usepackage{stackengine}
\usepackage{afterpage}
\usepackage{natbib}
\usepackage[affil-it]{authblk}

\newcolumntype{R}[2]{%
    >{\adjustbox{angle=#1,lap=\width-(#2)}\bgroup}%
    l%
    <{\egroup}%
}

\renewcommand{\arraystretch}{1.5}

\justifying

\title{Sails and Anchors: The Complementarity of Exploratory and Exploitative Scientists in Knowledge Creation \footnote{
The research leading to the results of this paper has received financial support from the French National Research Agency [reference: SEED -ANR-22-CE26-0013-01]. We also thank the fruitful exchanges with colleagues, especially: Pablo d'Este, Magda Fontana, Aldo Geuna, Jacques Mairesse, Julien Penin, Michele Pezzoni, Reinhilde Veugelers, Fabiana Visentin, Sandrine Wolff}}
\author[1,2]{Pierre Pelletier}
\author[2  \footnote{Email: \texttt{p.pelletier@unistra.fr; kevin.wirtz@unistra.fr}}]{Kevin Wirtz}
\affil[1]{\small 
CPS -- University of Turin, Italy}
\affil[2]{\small BETA -- University Strasbourg, France}

\date{}                 
\begin{document}
\onehalfspacing
\justifying

\maketitle
\begin{abstract} 
     This paper investigates the relationship between scientists' cognitive profile and their ability to generate innovative ideas and gain scientific recognition. We propose a novel author-level metric based on the semantic representation of researchers' past publications to measure cognitive diversity both at individual and team levels. Using PubMed Knowledge Graph (PKG), we analyze the impact of cognitive diversity on novelty, as measured by combinatorial novelty indicators and peer labels on Faculty Opinion. We assessed scientific impact through citations and disruption indicators. We show that the presence of exploratory individuals (i.e., cognitively diverse) is beneficial in generating distant knowledge combinations, but only when balanced by a significant proportion of exploitative individuals (i.e., cognitively specialized). Furthermore, teams with a high proportion of exploitative profiles tend to consolidate science, whereas those with a significant share of both profiles tend to disrupt it. Cognitive diversity between team members appears to be always beneficial to combining more distant knowledge. However, to maximize the relevance of these distant combinations of knowledge, maintaining a limited number of exploratory individuals is essential, as exploitative individuals must question and debate their novel perspectives. These specialized individuals are the most qualified to extract the full potential of novel ideas and integrate them within the existing scientific paradigm. 
\end{abstract}

\newpage

\section{Introduction}

This paper aims to study the extent to which the exploratory nature of scholars and the cognitive diversity of scientific teams shape their ability to generate innovative ideas and obtain scientific recognition. We propose a new author-level metric based on the semantic representation of past publications. It allows us to proxy both intra-individual and inter-individual cognitive diversities and their impacts on creativity in science.

A broadly accepted definition of creativity assumes a bipartite composition involving a combination of novelty and effectiveness \citep{runco2012standard}. Novelty, also called originality or invention, lies at the cornerstone of innovative research, bridging existing knowledge and unexplored scientific territories. Novelty serves as the foundation for peer recognition and functions as a ``reward system'' wherein the individual credited with the initial discovery garners recognition \citep{merton1957priorities,stephan1996economics,carayol2019}. Effectiveness, on the other hand, refers to the recognition attributed to this novelty. Novel research is frequently referred to as "High-Risk High-Reward" (HRHR) to reflect its high volatility of outcomes. Highly novel research receives more citations on average, but the uncertainty is also more considerable \citep{wang2017bias}. Due to its risky nature, funding opportunities are limited for innovative research \citep{ocde,franzoni2022funding}. The scientific system favours individuals who can ensure outcomes over those with potentially groundbreaking ideas that might disrupt the field \citep{alberts2014rescuing}. It has also been shown that a bias against novelty exists when scholars evaluate a peer's work \citep{wang2017bias,ayoubi2021does} and this effect is accentuated by the intellectual distance with the examiner \citep{boudreau2016looking}. Yet, these novelty indicators are still relatively recent and understudied as it is mostly intended to explain success or bias against novelty. As a result, it is essential to explore and validate new methods to better understand factors that promote novelty and its recognition.

Not all idea combinations are worth exploring, hence the challenge of distinguishing between novel and impactful ones. \cite{march1991exploration} supports the idea that a mix of exploitation and exploration is the key to an organization's survival. Transposed to science, exploration involves actively pursuing the expansion of one's understanding and curiosity across various areas of knowledge. Exploitation, on the other hand, refers to an individual specializing in a specific field and continuously building upon their expertise in that area. Put differently, producing a valuable invention would require a proper mix of typical and atypical combinations of knowledge, as seen in \cite{uzzi2013atypical}. This dichotomy has been studied in different domains, as mentioned in \cite{foster2015tradition} (e.g., "conformity" versus "dissent" in the philosophy of science), and can also be applied to research. The presence of individuals with exploratory profiles appears to facilitate communication among team members who are cognitively distant and then foster creativity \cite{page2007}. Indeed, people from outside a domain may have some advantage to offer fresh ideas through their distinct knowledge \citep{jeppesen2010marginality,kuhn1962structure}. 

As the body of knowledge in science expands, researchers increasingly specialize their competencies \citep{jones2008multi, jones2009burden} and thus are more able to recombine information locally in the knowledge space, facing incentives to collaborate \citep{fleming2001recombinant, boudreau2016looking}. Science is seen as a social phenomenon \citep{fleck2012genesis}. Thus, agents that recombine knowledge are individuals embedded in a social context, and cognitive and social phenomena strongly influence the invention process \citep{fleming2001recombinant}. Team size has been shown to impact creativity \citep{paulus2003group, shin2007educational, wuchty2007increasing, falk2011mapping, erren2017small, mueller2019building} and the cognitive dimension (i.e. differences in thinking, problem-solving approaches, and perspectives among individuals) plays a crucial role in enabling exchange of information and creation of new knowledge \citep{nooteboom2000learning,nooteboom2007optimal}. Cognitive diversity is shaped by individuals' characteristics and the trade-off between exploration and exploitation of the knowledge space during their careers. The cognitive distance among team members displays an inverted U-shaped correlation with both learning and innovation \citep{nooteboom2007optimal,ayoubi2017origins}, as people being too distant will face difficulty in communicating, and those being cognitively too similar can not benefit from distinct perspectives in the knowledge creation process.

Based on the concept of exploration and exploitation, we propose an indicator that serves as a proxy for exploratory \textit{vs.} exploitative trade-off at both the individual and team levels through past publications. In a nutshell, our indicator measures the cognitive distance between team members as well as the individual propensity to work on various subjects. This approach allows us to assess the impact of cognitive dimensions on individual and collective levels, acting as either a facilitator or a hindrance to creativity. We do not consider our indicator as a replacement for current novelty indicators but rather as a tool that could enhance our understanding of the mechanisms behind novelty and scientific impact. In fact, by incorporating the cognitive dimension into novelty studies, we can develop a more comprehensive understanding of the complex relationship between cognitive aspects, interdisciplinary efforts, and the nature of scientific innovation. Furthermore, examining these questions enables us to provide valuable insights and guidance for researchers and institutions striving to enhance scientific progress while avoiding potentially misleading interpretations of research performance measurement.

Using PubMed Knowledge Graph (PKG), we empirically investigate the role of these cognitive diversities in the production of novel research outcomes and the ability to obtain scientific recognition. We performed the analysis on novelty on five combinatorial novelty indicators \citep{uzzi2013atypical,lee2015creativity,foster2015tradition,wang2017bias,shibayama2021measuring}, both on references and MeSH terms, as well as on perceived novelty, using labels submitted by researchers to qualify the contribution of an article (Faculty Opinion)\footnote{More information can be found here: \href{https://facultyopinions.com/}{https://facultyopinions.com/}}. For scientific recognition, we rely on the traditional number of citations and six indicators of disruption and consolidation \citep{wu2019large,bu2019multi,bornmann1911disruption}.

Our findings emphasize the crucial role of cognitive dimensions in creativity, significantly impacting originality and success. We show that cognitive diversity always seems beneficial to combine more distant knowledge. In contrast, the within-team average exploratory profile follows an inverse U-shaped relation with combinatorial novelty (i.e. there is a turning point where it is no longer beneficial). The same relation can be found with citation counts, but we show that the cognitive dimension also strongly influences the nature of citations. Teams with more exploitative profiles consolidate science, while those with high exploratory profiles disrupt it only if they are associated with exploitative researchers. The union of those two types of individuals leads to the most disruptive and distant knowledge combinations. To maximize the relevance of these combinations, maintaining a limited number of highly exploratory individuals is essential, as highly specialized individuals must question and debate their novel perspectives. These specialized individuals are the most qualified to extract the full potential of novel ideas and integrate them within the existing scientific paradigm. This results can be linked to \cite{witt2009propositions}'s hypothesis 1, which states that the creation of new concepts involves two operations, one generative and one interpretative. The generative operation is carried out by the exploratory individuals, new ideas are then integrated into the existing scientific paradigm through a process of interpretation carried out by the exploitative individuals. The collaborative dynamic between exploratory and exploitative individuals facilitates both the generation of new concepts and their effective integration into existing theories.

The remainder of the paper is organized as follows. Section II features the background and literature review. Section III details the creation of our metrics and the methodology for addressing our research questions. Section IV presents the results of our analysis, and Section V concludes.

\section{Background and literature review}
\label{lit}

This section highlights the team's relevance in fostering creativity in science and emphasises how team size can influence this process. We also underscore the importance of identifying the social dimensions of the team, a crucial factor in generating new knowledge. Finally, we propose a new approach based on the semantic representation of authors' past publications that allows studying the role of the cognitive dimension in a team's ability to produce new and impactful knowledge.


 \subsection{Team science as an engine of creativity}
 
        Over the past two decades, there has been a significant increase in interest surrounding the Science of Team Science (SciTS) \citep{falk2011mapping}\footnote{For an up-to-date and comprehensive review, see \cite{wang2021science}.}. Since the 1950s, the average number of authors per paper has risen across all scientific disciplines \citep{wuchty2007increasing}. Research collaborations have also become more diverse, inter-institutional collaborations in science and engineering and social science grew by 32.8\% and 34.4\%, respectively, between 1975 an, d 2005 \citep{jones2008multi}. In addition, international collaboration has also expanded, with one in five research projects now involving multiple countries \citep{xie2012american}.

        Teamwork has proven to be a practical approach to producing impactful scientific results. Articles written by teams tend to have a higher impact, receiving more citations on average and are more likely to become influential than articles authored solely \citep{wuchty2007increasing,whitfield2008group}. Researchers benefit from collaboration in various ways. Collaborative efforts can enhance rigour through co-authors' verification \citep{leahey2016sole} and facilitate the dissemination of their work beyond their immediate networks \citep{leahey2016sole}; this effect is further amplified when collaborations are international or inter-institutional \citep{adams2013fourth,jones2008multi}. Additionally, teams have better access to resources, as projects executed by groups are more likely to apply for funding and succeed in obtaining it \citep{rawlings2011influence}. Teams are more likely to produce novel articles than solo-authored publications \citep{carayol2019, uzzi2013atypical, wagner2019international}. As highly cited work is often associated with a combination of novel and conventional ideas \citep{uzzi2013atypical}, teams of researchers may be more adept at generating novel ideas or striking a balance between novel and traditional concepts than individual authors.

        Successful team performances put individuals and their interactions at the heart of the creative process.
         Over recent decades, the perception of teamwork has undergone significant changes. In the early 1990s, the prevailing belief was that groups should not be used for creativity because of inherent process loss in the creative process. This perspective has shifted dramatically, and team collaboration is now considered a critical factor in promoting creativity \citep{paulus2003group}. Creativity relies on individual's existing knowledge base: ``\textit{Creative thinking cannot happen unless the thinker already possesses knowledge of a rich and/or well-structured kind}'' \citep{boden2001creativity}. Knowledge exists on a continuum, ranging from explicit to tacit \citep{nonaka1994dynamic}. The generation of new knowledge occurs through interactions between explicit and tacit knowledge via a process known as the socialization, externalization, combination, and internalization (SECI) spiral. \cite{tahamtan2018creativity} highlighted various approaches reported by researchers for fostering creativity. Engaging in conversations with colleagues seems to remain central to problem-solving and generating new, practical ideas. Hence, new ideas are becoming more challenging to discover as the idea space expands linearly while scientific publications grow exponentially \citep{bloom2020ideas,milojevic2015quantifying}. As scientific knowledge increases, team sizes grow, and agents increasingly specialize their competencies \citep{jones2008multi,jones2009burden}. 
         
         The burst of possible combinations in the knowledge space suggests that agents can more effectively recombine information locally \citep{fleming2001recombinant}. ``Local search'' for an inventor involves exploiting existing combinations or using standard technological components. Agents tend to direct their research towards familiar subjects, focusing on topics related to their expertise or that of their co-authors (local search/exploitation) \citep{fleming2001recombinant,nelson1985evolutionary,march1991exploration}. Conversely, exploration (or "distant search") is characterized by using new components or testing novel combinations \citep{fleming2001recombinant,march1991exploration}. The nature of the new combinations realized depends on agents' trade-offs between exploiting and exploring the knowledge landscape. Exploitation reduces the risk of failure, as researchers draw from experience with combinations and architectures that have previously failed \citep{vincenti1990engineers}. Researchers must then collaborate with others to explore the knowledge space more efficiently, and the team's composition might determine this balance between exploration and exploitation.
        
 \subsection{Team characteristics in the creative process}
        We review here some dimensions of the team composition that affect the scientific process.
        
         \underline{\textit{Size dimension}}: The importance of co-authors during the process of creativity has been debated in the literature, and the effect of team size and composition on creativity has been the focus of multiple studies \citep{paulus2003group, shin2007educational, wuchty2007increasing, falk2011mapping, erren2017small,mueller2019building}. Team size shapes and is shaped by the nature of the work carried out. Large teams tend to be more risk-averse and consolidate a field rather than introducing new opportunities \citep{christensen2003innovator,paulus2013understanding,lakhani2013prize,wu2019large}. Larger teams use more up-to-date and influential research in their work, consequently fostering greater engagement within their scientific community and further increasing their impact \citep{wu2019large}. However, large teams are more prone to coordination and communication failures as the entire team must have faith in the project to succeed, as agreement and communication between team members can be challenging and time-consuming \citep{bikard2015exploring}. In fact, the number of people involved in a project can have heterogeneous effects on creativity, and no optimal team size fits every project. A small team may be more useful in the conceptualization phase, while a larger team might be beneficial in the implementation and testing phase of the project \citep{wang2021science}. \cite{shin2007educational} highlight the organization's importance for creativity. Using evidence from Cambridge and AT\&T's Bell Laboratories (home to numerous Nobel Prize winners), they discuss researchers' ideal context for fostering creativity and conclude that the presence of a healthy environment for a small group of people (up to seven) promotes creativity. These results are further confirmed by \cite{lee2015creativity} and \cite{carayol2019}, indicating that the relationship between team size and novelty appears U-shaped and is highly heterogeneous across disciplines. \\

        \underline{\textit{Structural and relational social capital}}: \cite{nahapiet1998social} conceptualize three dimensions of social capital that impact intellectual capital development: structural, relational and cognitive. 
        Though primarily used to understand intellectual capital development in organizations and firms, the dimensions of social capital presented in \cite{nahapiet1998social} can be applied to the context of knowledge production in science due to their intrinsic relevance to relationship and network dynamics \citep{liao2011improve}. Structural capital examines the links between individuals, and structural distances have been widely studied through collaboration networks (see \cite{kumar2015co} for an extensive review on network collaborations). Relational capital represents the nature and intensity of the connections between team members. A critical factor in intellectual development is the ability to communicate with each other, and the actors' experience reinforces the phenomena \citep{taylor2006superman, liao2011improve, kelchtermans2020off}. For instance, \cite{mcfadyen2004social} emphasize the role of the intensity of past relationships between scientists in fostering new knowledge. Indeed, members with strong relationships, norms, obligations, and mutual trust tend to communicate more easily \citep{liao2011improve}. Other relational aspects, such as hierarchical or geographical dimensions, also impact the knowledge space exploration. For example, supervising doctoral students is not only associated with entering new areas but also extending towards more distant fields \citep{kelchtermans2020off}\\

        \underline{\textit{Cognitive social capital}}: The cognitive capital remains challenging to measure as it is linked to the shared background between coauthors and their common language. Cognitive diversity is often encouraged through interdisciplinary projects as the intersection of different perspectives is commonly required to solve complex scientific problems \citep{page2007}. Indeed, people from outside a domain may have some advantage to offer fresh ideas through their distinct knowledge \citep{jeppesen2010marginality,kuhn1962structure}. The effectiveness of generating new knowledge is impacted by factors such as variations in background, belief and reasoning styles among scientists, all of which contribute to cognitive diversity. The cognitive distance between team members is expected to display an inverted U-shaped correlation with both learning and innovation \citep{nooteboom2007optimal}, as people being too distant will face difficulty in communicating, and those being cognitively too similar benefit less from distinct perspectives in the knowledge creation process.
        
        Cognitive distances between individuals can be studied through various metrics. \cite{kumar2017author} used, for example, citations networks and citations context in full text. 
        \cite{boudreau2016looking} represented the cognitive distance between funding evaluators and the proposal through MeSH terms similarity. Similarly, \cite{ayoubi2017origins} represent the distance between the focal scientist and her team by comparing cosine similarities of referenced journals from scientists' past publications. Other measurements, without being explicit, may relate to cognitive dimensions, \cite{wagner2019international} discovered that international collaborations negatively affect novelty and produce more conventional knowledge combinations, highlighting barriers and transaction costs that influence the production of creative work.
        Finally, measures of cognitive distance strongly relate to interdisciplinarity. \cite{petersen2021grand} represent author diversity using the discipline of the institution. \cite{bramoulle2018length} calculate team specialization using the Herfindahl index, derived from the proportion of past publications across different fields as indicated by the first digit of JEL codes. They find this specialization negatively impacts combinatorial novelty. Using authors' disciplinary diversity, \cite{abramo2018comparison} show that more distant coauthors produce articles with more diverse references. \\

        \underline{\textit{Exploratory profile}}: Individual characteristics and the ability to interact with individuals from different fields are essential to efficiently managing cognitive diversity in a team. When the distance between disciplines is too high, a ``Renaissance'' individual \citep{jones2009burden} can ease their connection \citep{wu2022metrics}. The presence of a scientist with a multifaceted profile bridges the gap between the different backgrounds of other team members. This is crucial as a shared knowledge base between researchers streamlines the socialization process and facilitates knowledge recombination, fostering creativity. \cite{shin2007educational} focused on the relationship between diversity (interdisciplinarity) and creative ideas in groups. \cite{shin2007educational}'s idea is that the presence of a "transformational leader", whose role is to mediate between individuals, each specialized in a different field, leads to greater team creativity. \cite{xu2022flat} provided a first answer to this hypothesis by examining the share of team members engaged in the conceptual work, the L-ratio, which was deduced from the analysis of author contribution reports. The findings suggest that hierarchical teams generate less novelty than egalitarian teams and tend to develop existing ideas more frequently.\footnote{\justifying Interestingly, their method was expanded in an article with no contribution reports. Through Louvain algorithms, they identified clusters of co-occurring research activities in their first dataset. They then built a neural network to infer author roles based on their characteristics and predicted it for 16 million articles on Microsoft Academic Graph (MAG).} We argue that the notion of transformational leader or renaissance individual is connected to exploratory profile \textit{à la} \cite{march1991exploration}, individuals enabled to link others in the knowledge space due to their ability to navigate in different spaces.

 \subsection{Exploring the cognitive dimension}

        We investigate scientific impact through citation networks and recent indicators of disruption and breath and depth \cite{wu2019large,wu_wu_2019,bu2019multi,bornmann1911disruption}. These indicators determine whether a document consolidates a domain or constitutes a founding step. To explore its influence on novelty, we use two approaches, one based on combinatorial novelty indicators \citep{uzzi2013atypical,lee2015creativity,foster2015tradition,wang2017bias,shibayama2021measuring} and one based on external validation via Faculty Opinion (previously called F1000) following \cite{bornmann2019do}. Faculty Opinion is a website hosting reviews of papers tagged as presenting ``New Results'', ``Novel Drug target'', ``Technical advancement'', ``Interesting hypothesis'', and ``Controversial results'', among other categorizations labelled by experts in the field. It allows us to empirically assess the capacity of novelty indicators and our indicators to predict the novelty as perceived by other researchers in the community. 
        
        Novelty indicators have been compared and evaluated based on citation count \citep{uzzi2013atypical, lee2015creativity, foster2015tradition, wang2017bias}. \cite{fontana2020new} compared \cite{wang2017bias} and \cite{uzzi2013atypical,lee2015creativity} using randomized citation networks and demonstrated the ability of the \cite{uzzi2013atypical,lee2015creativity} indicators to better track novelty. Their findings are supported by using some Nobel Prize winners' articles and a list of APS milestone articles. Other studies have evaluated these indicators based on surveys, such as \cite{shibayama2021measuring} and \cite{matsumoto2021introducing}, whereas \cite{bornmann2019do} have evaluated them based on labels collected on Faculty Opinion and found similar results as in \cite{fontana2020new}. However, only a few indicators have been compared and tested simultaneously. This study intends to validate the effect of the cognitive dimension on a large variety of metrics.
      
        Our indicator is not a substitute for other novelty indicators. It does not represent the novelty of an article as it is based upon previous information and would be similar even without the focal article. Instead, it provides an understanding of team composition that would benefit creativity in science.
        We can think of our measure as a measure of \textit{potential novelty}, i.e. opportunities for new knowledge recombination available through the diversity of background in the team and the capacity of individuals to bridge the gap between other team members. In comparison, combinatorial novelty indicators would capture then the \textit{realized novelty}, i.e. the output of the research conducted by this team in terms of pieces of knowledge used. Finally, Faculty Opinion labelling and other external validation methods can describe the \textit{perceived novelty}, i.e. the peers' perception of this study. Hence, in these terms, we ask whether potential novelty contributes to realized and perceived novelty and its scientific recognition. Two research questions can be drawn regarding the effect of the cognitive dimension on creativity. Do teams with higher cognitive diversity are more likely to approach a subject creatively, demonstrating originality (\textit{perceived} and \textit{realized}) and recognition? Does the presence of \textit{exploratory} individuals within a team enhance communication among members and facilitate their exploration of the knowledge space to develop new and relevant solutions to research problems? Studying the cognitive dimension of creativity in science is of great interest, especially as it can help identify how to improve collaboration and communication among researchers with diverse cognitive profiles. Through our metric, we also offer a different approach to resource allocation decisions, giving another picture of teams with a high potential for creative output. \\

\section{Data and methods}\label{3}

\subsection{Measuring cognitive diversity and exploratory profile}\label{3.1}

        The proposed metric examines the semantic heterogeneity of researchers' work as a proxy for their cognitive diversity. It thus offers an alternative to using categories, keywords, or citation networks, more complex to be monitored directly by the researchers themselves. Following \cite{hain2020text} and \citep{shibayama2021measuring}, we can embed this list of documents in a vectorial space to apply a distance measure such as cosine similarity \citep{mikolov2013distributed}. We assume that an author of a paper in a specific position within the semantic space possesses knowledge embedded around that position. Our indicator has two properties: it offers a measure of researchers' profiles at the individual level and a measure of distances between them. Consequently, we can proxy the trade-off between exploitation and exploration that a researcher undergoes throughout their career (intra-individual) and the trade-off materializing during the formation of a team (inter-individual) within the same mathematical space. 
    \begin{figure}[h!]
      \centering
      \includegraphics[width=\textwidth]{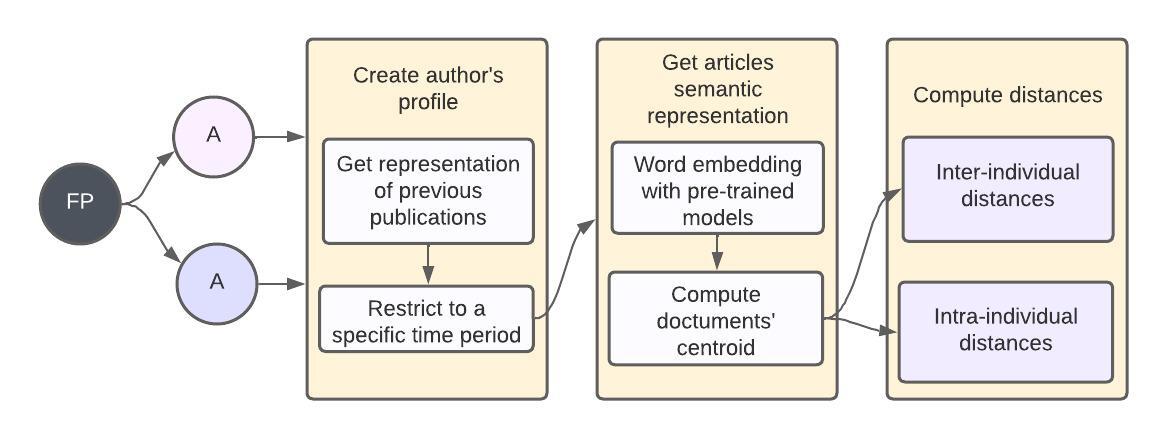}
      \caption{Construction of the indicator}
      \label{figure:mapidic}
    \end{figure}

    As explained in Figure \ref{figure:mapidic}, we track authors to create a list of authors' past publications. Then, we can create a cognitive profile for each author at a given time $t$; each publication is embedded in the semantic space and represents the cognitive landscape of the author. We restrict to publications up to $b$ years before $t$ to account for researchers' current topics of interest and difficulty retaining information \citep{argote1990persistence}. We can finally define a researcher's exploratory profile at time $t$ by calculating pairs of cosine distances between past papers published. This will create a density of cosine distances which, using the taxonomy of \cite{march1991exploration}, can be interpreted the following way: the fatter the right (left) tail is, the more exploratory (exploitative) the researcher. The same holds for the team. A sizeable right tail indicates a cognitively distant researchers team. This provides us with information on how distant their knowledge base is from others. Note that our measure implicitly incorporate the fact that cognitive distance between two researchers decreases with when frequency of interaction increase, as more articles will be similar when calculating the distribution of distance between the two researchers. An intra-author and inter-author distribution enables a wide exploration of the relationship between novelty, creativity, and teams.

    \begin{figure}[h!]
      \centering
      \includegraphics[width=\textwidth]{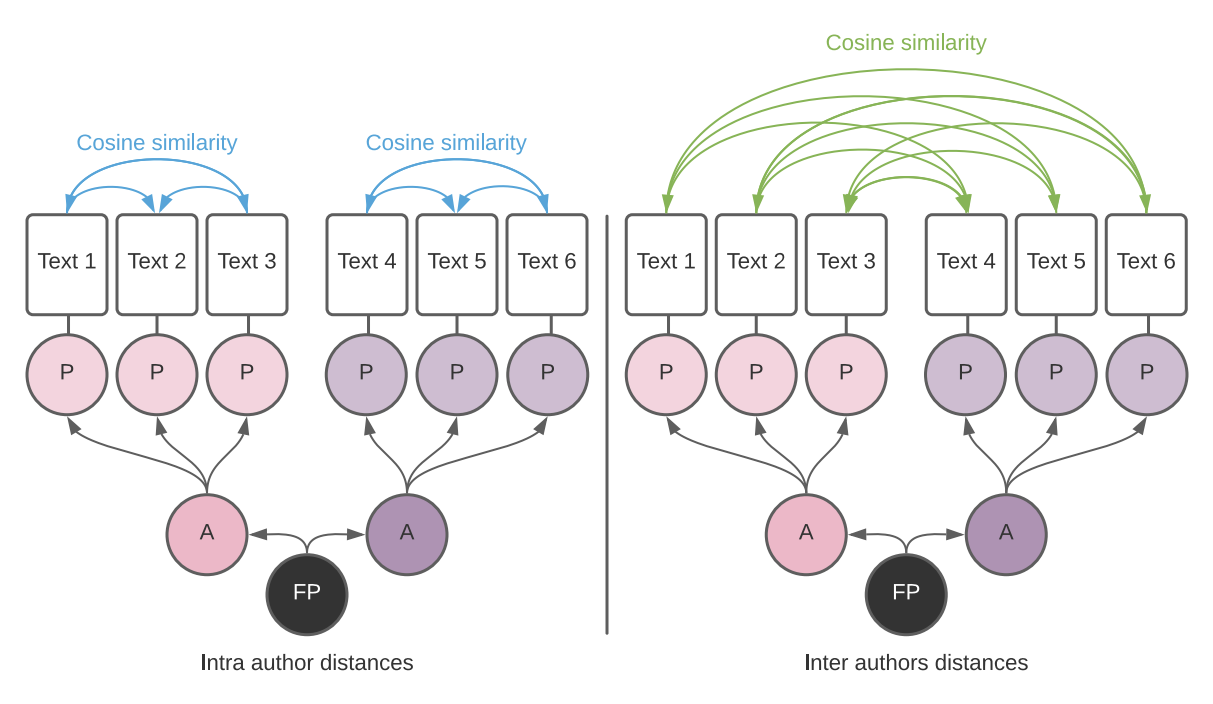}
      \caption[Exploratory profile and cognitive diversity]{Exploratory profile and cognitive diversity}
      \label{figure:indicators}
    \end{figure} 
    
        We model our measures on two different perspectives as represented in Figure \ref{figure:indicators}: intra-author distances, which asses exploratory profile, and inter-authors distances, to capture cognitive diversity. A given focal paper (FP) is written by two authors named 'A'. We retrieve each author's past production, named 'P'. On the one hand, we can then calculate the distance between all publications from a given author (intra-author distances). On the other hand, we can also compare past publications from two authors (inter-author distances). We can build our framework using directed bipartite networks, defined as $G(U,V,E)$. $U$ represents the set nodes for authors, $V$ is the set for articles, and $E$ is the set of links between authors and articles.        
    
        We only consider collaborations between authors when looking at a given article; collaboration is implicit since the set of parents of a given document $FP \in V$ corresponds to the set of authors that collaborate. For a given document $FP$, an author that has contributed to $FP$ is noted $a$, and the set of nodes that contributes to $FP$ is then $In_{FP} = \{a\in U : (a,FP) \in E\}$. 
        
        We want to retrieve all past publications for all authors in $In_{FP}$. The global set of publications before $FP$ is noted $V^{t-b}_{FP}$, the set of articles published $b$ years before the document $FP$. The set of past publications for author $a \in In_{FP}$ is noted $Out^{t-b}_a = \{v \in V^{t-b}_{FP} : (a,v) \in E \}$. For a document $FP$ and an author $a \in In_{FP}$, we retrieve the set of past publications $Out^{t-b}_a$. All $Out_a^{t-b}$ elements vectorial representations are compared, from which distribution of cosine distances are calculated. The distance between two documents $i, j \in Out^{t-b}_a$ is $d_{ij} = 1-COS(T_i,T_j)$ where $T_i$ is the dense vector text representation for document $i$.

        \vspace{0.5cm}
        \noindent\textit{\underline{Intra-author semantic distances}}: A distribution of semantic distance score $D_a$ is computed through cosine similarity using all document  $i, j \in Out_a^{t-b}$, the process is repeated for each authors $a \in In_{FP}$. The intra-author distance for a given author $a$ is the q-th percentile ($P_q$) of this distribution and is written as:
        
        $$Intra_{a} = P_{q}(D_a)$$
        
        A general distribution of the intra-authors publication distances is constructed using the set of distances for all authors $A_{FP} = \{D_a: a \in In_{FP}\}$, the individual trade-off between exploitation/exploration is then captured through the average of the exploratory profiles in a given team.
        
        $$Intra_{FP} = \frac{\sum_a (P_{q}(D_a))}{|In_{FP}|}$$

        \vspace{0.5cm}
        \noindent\textit{\underline{Inter-authors semantic distances}}: A distribution of semantic distance score between authors' previous work is constructed by comparing different authors' publications. For two given authors $a, e \in In_{FP}$, $|Out_a^{t-b}|\times|Out_e^{t-b}|$ distances are used to construct the distribution of distances $D_{a,e}$ between $a$ and $e$. The final distribution then groups together all distances between authors' previous works $B_{FP} = \{D_{a,e}: a, e \in In_{FP}\}$, the trade-off between exploitation/exploration in team composition is captured through the percentile of $B_{FP}$:
        
        $$Inter_{FP} = P_{q}(B_{FP})$$

        Current techniques for large-scale author disambiguation allow the investigation of individual trajectories in science. However, the use of this information comes with a computational cost.
        This indicator pushes towards a massive use of data because one needs all authors' past publications for a given set of documents. Structuring the data to compute the measure is time-consuming and data-intensive. One needs indeed all papers' text from all authors in a given database.
        However, using pre-trained embedding models allows direct computing indicators without the requirement of complete database access. Therefore, measures are not dependent on the study sample as indicators of novelty based on cooccurrence matrices but rather on the sample used to train the model. Also, by processing titles and abstracts through embedding techniques, the authors' background is represented with greater granularity than through the keywords or the journals where the authors have been published.

\subsection{Data}

        Our analysis relies on two databases. The first, PubMed Knowledge Graph (PKG), allows us to test the effect of the cognitive dimension on scientific impact and \textit{realized} novelty of articles, while the second, Faculty Opinion verifies whether the cognitive dimension affects the \textit{perceived} novelty by peers. 
            
       We use Pubmed Knowledge Graph (PKG), a collection of 35 million scientific papers and books from life science and biomedical journals provided by the National Library of Medicine (NLM) at the National Institutes of Health (NIH). Authors are disambiguated by leveraging Natural Language Processing (NLP) and online data, as outlined by \cite{xu2020building}. We based our analysis on all the 3.5M articles written by 3,276,250 authors and published in 9,348 journals between 2000 and 2005. We selected fairly old data due to the nature of the process studied. Indeed, novel articles are more likely to become "sleeping beauties" and accumulate citations in the long run \citep{lin2021novelty}. Also to compute novelty indicators, we require information about references. We rely both on abstracts of references to embed their semantics and calculate the distance as in \cite{shibayama2021measuring}. Also, we use past publication references' journals to build past cooccurrence matrices used to capture combination existence and difficulty for other novelty indicators. For this purpose, we used the database between 1980 and 2005 to get all information needed, representing 11,261,955 documents.
        
        To test if our indicators affect the novelty perceived by peers, we used Faculty Opinion following \cite{bornmann2019do}. Faculty Opinion is a database featuring papers tagged as presenting 'New Results', 'Novel Drug target', 'Technical advancement', 'Interesting hypothesis', and 'Controversial results', among other categorizations determined by the platform users. The platform hosts reviews of the most significant research in Biology and Medicine. This makes it easy to match the articles in the database with PKG. Indeed, from the 190k articles in Faculty Opinion, we found 27,122 in our sample (2000-2005).

\subsection{Empirical strategy}

        To explore the relationship between the team's cognitive dimension and its ability to recombine pieces of knowledge in novel ways and achieve recognition, we start with a basic exploratory data analysis followed by three econometric analyses to test our hypotheses.
        
        The first two analyses aim to understand how a team's cognitive diversity and the exploratory profiles of its members impact \textit{perceived} novelty (i.e., peer labelling on Faculty Opinion) and \textit{realized} novelty (i.e., indicators of combinatorial novelty). Then our analysis seeks to comprehend the effect of the cognitive dimension on scientific recognition using citation and disruption measures.
        
        \textit{Realized} novelty and scientific impact connections with cognitive dimension are both investigated through PKG, the normalization performed at the field and year levels of this measure provides a measure ranging between 0 and 1, which we model using linear models with cluster robust standard errors at the journal level. Lastly, we examine how the presence of highly exploratory and exploitative individuals influences the team's creativity. This analysis will help determine if cognitive diversity and the presence of exploratory profiles are explicitly visible in an article's knowledge composition.

        For the analysis of \textit{perceived} novelty, we employ the Faculty Opinion database and model, through Logit and Poisson regressions, the likelihood of an article being labelled with ``novel'' categories (``Technical Advance'', ``Interesting Hypothesis'', ``Novel Drug Target''). In our sample, 80\% of the observations are labelled as 'New Findings', and 95\% of the total sample would be considered new using the top 4 most represented categories (22,216 novel articles versus 1,750 not-novel). The fact that most articles are labelled as new findings makes this category less informative; therefore, we decided to exclude it and remove articles solely labelled with this category. As a result, our prediction is based on a more balanced sample (8,950 novel articles versus 3,605 not-novel). This will enable us to understand whether the cognitive dimension is associated with \textit{perceived} novelty. We do not expect a direct effect but rather hypothesize that cognitive diversity influences a latent variable representing the article's actual contribution. This actual contribution of the paper may or may not be visible in the \textit{realized} novelty measured by novelty indicators but might be then reflected in labelling made by peers.

\subsection{Variables}

Variables used in our empirical analysis can be separated into four categories: novelty indicators, scientific impact, cognitive, and control variables. For control variables, aside from data from PKG, we use journals listed in Scimago to control for scientific domains and measure of the impact associated with the journal. Each of our variables is at the paper level. For the empirical strategy, novelty, impact and cognitive measures will be field weighted by year using the percentile rank procedure -- noted (FW). We use the first category of the journal from Scimago to approximate the field. 

\subsubsection*{Novelty indicators}
The indicators used in our analysis are \cite{uzzi2013atypical}, \cite{lee2015creativity}, \cite{foster2015tradition}, \cite{wang2017bias}, \cite{shibayama2021measuring}. A formal mathematical description of them can be found in \cite{pelletier2022novelpy}. Note that we have inversed the sign of the measures related to \cite{uzzi2013atypical} for simplicity and comparison with other indicators. The computation is done with \textit{Novelpy}, a python package that allows computing novelty and disruptiveness indicators.\footnote{\justifying See \href{https://novelpy.readthedocs.io/}{https://novelpy.readthedocs.io/}}.

\subsubsection*{Scientific impact variables}
For impact measures, we use citation counts and disruptiveness indicators, also described in \cite{pelletier2022novelpy}. We used all available indicators in \textit{Novelpy}, namely: \cite{wu2019large}, \cite{bu2019multi} and \cite{bornmann1911disruption}.

\subsubsection*{Cognitive variables}

\noindent\textit{\underline{Team cognitive diversity:}}
The mean of the inter-authors semantic distance as defined in Section \ref{3.1} with q=90 for a given paper. It measures to what extent a team is composed of highly cognitively distant authors (i.e. Author 1 background is vastly dissimilar to Author 2 background). Furthermore, we suppose the relation between the team's cognitive diversity and other measures is not linear. We take the square of the team's cognitive diversity to test this.
\newline

\noindent\textit{\underline{Average exploratory profile:}}
The mean of the intra-authors semantic distance as defined in Section \ref{3.1} with q=90 for a given paper. It captures to what extent a team comprises authors with distant past publications (i.e. Author 1 worked on diverse subjects). As for team cognitive diversity, we add a square term in the regressions.
\newline

\noindent\textit{\underline{Number of highly exploratory authors:}}
To have more information on the team structure, we decided to define a threshold to identify highly exploratory authors. Looking at the intra- author's semantic distance as defined in Section \ref{3.1}. An author is considered highly exploratory if its 90$^th$ percentile is in the top 10\% of all $Intra_{FP}$ in our sample. \newline

\noindent\textit{\underline{Number of highly exploitative authors:}}
We expect highly exploratory authors to work best with highly exploitative authors (i.e. Novelty is probably most successful with a combination of typical and atypical individuals). We construct this measure following the same procedure as exploratory authors. Looking at the intra- author's semantic distance as defined in Section \ref{3.1}. An author is considered highly exploitative if its 90$^th$ percentile is below our sample's median of all $Intra_{FP}$. \newline

\noindent\textit{\underline{Interaction term between highly exploratory and  highly exploitative authors:}}
We added an interaction term between the two types of profiles to understand how they complement each other.

\subsubsection*{Control variables}

We included as control variables the number of authors, references and MeSH terms. We also controlled for the year and information related to the journal of publication.\newline

\noindent\textit{\underline{Scimago Journal Ranking (SJR):}}
An indicator of a journal's prestige based on weighted citation and eigenvector centrality derived from Scopus' citation networks by Scimago \citep{gonzalez2009sjr}. \newline

\noindent\textit{\underline{Scimago Journal Category:}}
Scimago provides a classification of journals based on various fields. We used the first category linked to a journal; our database contains journals from 271 categories. 

\subsection{Descriptive statistics and preliminary evidence}
        
We further clean our database and restrict it to papers with at least 2 references/ MeSHterms/ authors and with a journal ISSN. Our final dataset represents approximately 2.1M articles. 

Table \ref{tab:desc} presents the descriptive statistics for the variables in our sample. Examining their distribution, it is worth noting that some indicators concentrate novelty around a small number of articles, as in \cite{foster2015tradition} or in \cite{wang2017bias}, merely 21\% of the articles possess non-zero values (measured on references). Also, indicators such as citation count or \cite{uzzi2013atypical} among others, display relatively extreme values. Specifically for \cite{uzzi2013atypical}, it is highly dependent on the z-score computation, when the variance of the journal combination is minimal, the z-score can rapidly become substantial. These disparities in distribution prompted us to apply a percentile rank procedure by field and year, as explained in the previous subsection.

\begin{table}[h] \centering 
\renewcommand{\arraystretch}{1}
\setlength{\tabcolsep}{0.15pt}
  \caption{Descriptive statistics} 
  \label{tab:desc} 
    \scalebox{1}{
	\begin{threeparttable}
\begin{tabular}{@{\extracolsep{5pt}}lcccccccc} 

\midrule
\midrule

Statistic & \multicolumn{1}{c}{Min.} & \multicolumn{1}{c}{Pctl(25)} & \multicolumn{1}{c}{Median} & \multicolumn{1}{c}{Mean} & \multicolumn{1}{c}{Pctl(75)} & \multicolumn{1}{c}{Max} & \multicolumn{1}{c}{St. Dev.} & \multicolumn{1}{c}{N} \\ 

\midrule

\# References &  2 &   12 &   22 &  27.37 & 36 & 2690 & 25.76 & 2108280\\
\# Meshterms &2 &  9 & 13 & 13.25 & 16 & 51 & 5.19 & 2108280\\
\# Authors  & 2 &   3 &   4 &  5 & 6 & 282 & 2.94 & 2108280\\
\# Citations &   0 &     9 &    22 &   46.99 &  50 & 81577 & 129.47 & 2108280\\
SJR &0.1 &  0.627 &  1.130 & 1.787 &2.035 & 39.946 & 2.22 & 2094669\\
\addlinespace
Disruption$_{1}$  & -1 & -0.007 & -0.001 & 0.003 &05179 &  1 & 0.06 & 2108280\\
Disruption$_{1noK}$ & -1 & -0.588 & -0.269 & -0.192 &0.111 &  1 & 0.51 & 2108280\\
Disruption$_{5}$  & -1 &  0 &  0.001 & 0.018 &0.009 &  1 & 0.07 & 2108280\\
Disruption$_{DeIn}$ &0 &  0.79 &  1.662 & 2.067 &2.875 & 92.5 & 1.81 & 2108280\\
Breadth & 0 & 0.307 & 0.5 & 0.517 & 0.714 & 1 & 0.26 & 2108280\\
Depth & 0 & 0.258 & 0.5 & 0.458 & 0.672 & 1 & 0.26 & 2108280\\
\addlinespace
Share Exploratory & 0 & 0 & 0 & 0.063 & 0 & 1.0 & 0.14 & 2108280\\
Share Exploitative & 0 & 0 & 0.333 & 0.365 & 0.6 & 1 & 0.32 & 2108280\\
Author intra $_{abs}$  & 0 & 0.22 & 0.29 & 0.29 & 0.36 & 1.02 & 0.09 & 1837749\\
Author inter $_{abs}$  & 0 & 0.26 & 0.33 & 0.33 & 0.40 & 1.02 & 0.09 & 1837748\\
\addlinespace
Shibayama $_{abs}$ & 0 & 0.222 & 0.274 & 0.275 & 0.327 & 0.991 & 0.07 & 2081854\\
Uzzi$_{Ref}$   & -62396.32 &     -7.34 &      3.66 &   -18.03 &   14.02 &    199.49 & 206.82 & 1891079\\
Lee$_{Ref}$ & -17.581 &   0.145 &   0.840 &  0.567 & 1.466 &   6.006 & 1.45 & 2092283\\
Foster$_{Ref}$  & 0 & 0.117 & 0.4 & 0.366 & 0.583 & 1 & 0.25 & 2092283\\
Wang$_{Ref}$&  0 &    0 &    0 &   0.583 &  0 & 2872.106 & 4.79 & 2092283\\
Uzzi$_{Mesh}$ & -287.0 &   -1.1 &    0.9 &   2.7 &  4.5 &  189.1 & 8.19 & 765751\\
Lee$_{Mesh}$   & -7.996 &  0.4562 &  0.807 & 0.794 &1.174 &  4.717 & 0.60 & 2105186\\
Foster$_{Mesh}$    & 0 & 0.274 & 0.476 & 0.424 & 0.591 & 1 & 0.22 & 2105186\\
Wang$_{Mesh}$  &0 &  0 &  0 & 0.299 &0.307 & 28.668 & 0.76 & 2105186\\
\midrule
\bottomrule
\end{tabular} 
 \end{threeparttable}
 }
\end{table}

\begin{figure}[h!]
  \centering
  \includegraphics[width=0.7\textwidth]{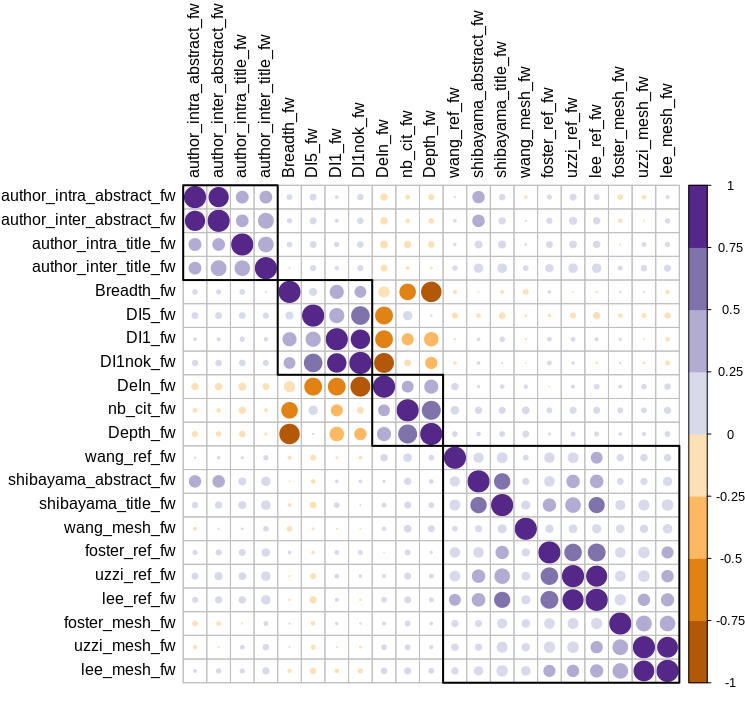}
  \caption{Correlogram with hierarchical clustering}
  \label{figure:correlogram}
\end{figure}

The correlogram in Figure \ref{figure:correlogram} illustrates the various indicators' interconnection. A hierarchical clustering algorithm is applied to the correlation matrix and several clusters emerge. It includes citation and consolidation indicators, novelty indicators, cognitive dimension indicators, and disruption indicators. Regardless of whether MeSH terms or references are used to derive the indicators, the novelty indicators group remains consistent, suggesting that combinatorial novelty indicators capture a shared underlying dimension of innovation in scientific research. The correlation between \cite{lee2015creativity} and \cite{uzzi2013atypical} is particularly robust since both measures are nearly identical except for the incorporation of the reference's publication year in \cite{uzzi2013atypical}'s resampling process. It should be noted that a negative correlation is expected since low values signify atypicality in \cite{uzzi2013atypical}, while high values represent novelty in \cite{lee2015creativity}, this is why we inverse the sign of \cite{uzzi2013atypical} to get positive correlation between indicators. A strong correlation is observed between \cite{shibayama2021measuring} and our indicators, as it employs the same measurement on references, and some elements may overlap. Specifically, self-citation increases the correlations between \cite{shibayama2021measuring} and our indicator since the same combinations are calculated in the author and reference parts. Moreover, the clustering differentiates between citation count, consolidation indicators (Depth, DeIN), and disruption indicators (DI1, DI5, DI1nok, and Breadth). These distinctions emphasize how consolidation indicators are more closely related to citation count and demonstrate how disruption indicators capture other dimensions of scientific impact.

\begin{figure}[h!]
\centering
\begin{subfigure}{.52\textwidth}
  \centering
      \includegraphics[width=\textwidth]{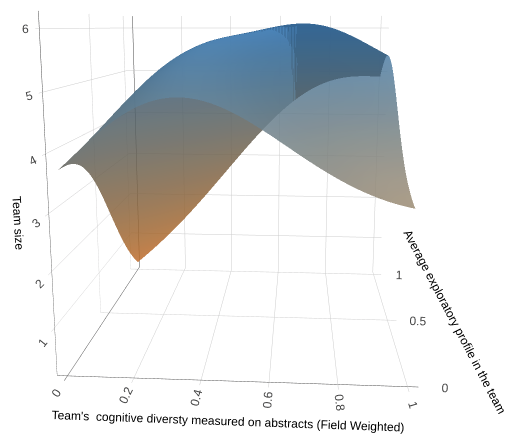}
\end{subfigure}%
\begin{subfigure}{.52\textwidth}
      \includegraphics[width=\textwidth]{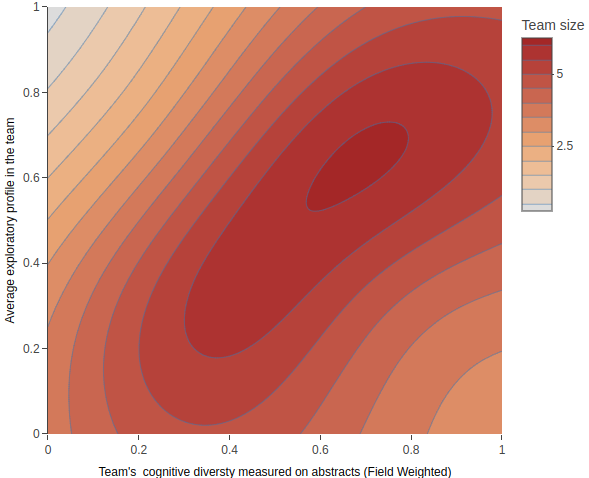}
\end{subfigure}
\caption{Team size, exploratory profiles and cognitive diversity}
      \label{figure:team_3d}
\end{figure}

The development of an author-level indicator necessitates examining its relationship with team size. Figure \ref{figure:team_3d} illustrates how intra- and inter-individual cognitive indicators are strongly associated with team size. Although it is unclear whether cognitive diversity generates a specific team size or if team size produces this diversity, it is visible that as the cognitive diversity within a team increases, the average exploratory profile must also rise to maintain a comparable team size. The hump-shape relationship on both sides is easily observable, suggesting that the more diverse the team and/or the more exploratory the individuals, the smaller the team. Conversely, highly homogeneous teams typically imply smaller average team sizes, even if the average exploratory profile is high. This pattern is partially attributable to the construction of our indicator, which averages distance. In larger teams high distance between members might be compensated by other members that are close to each other. This counterbalancing is less pronounced in smaller teams, resulting in more extreme values. However, several explanations for this phenomenon can be offered. For instance, substantial cognitive diversity might create communication barriers among team members, particularly when individuals are less exploratory. Consequently, smaller teams are formed due to potential coordination and knowledge exchange difficulties. In cases with a high average exploratory profile combined with high cognitive diversity, forming smaller teams may be more convenient, as researchers might explore the knowledge space too broadly. Smaller teams could help prevent efforts from dispersing in various directions. As for teams with low cognitive diversity, the absence of cognitive diversity and exploratory profiles could relate to niches where individuals possess similar knowledge and expertise. As a result, many team members might not be necessary, as they can efficiently navigate the local knowledge space. The same argument can be made for individuals with comparable skills and exploratory profiles, as they may represent teams that regularly collaborate on diverse topics. The distinct skill requirements for these teams may be lower, leading to smaller team sizes.

Interestingly, when comparing the analysis of team size with disruption, we confirm the findings of \cite{wu_wu_2019}. As illustrated in Figure \ref{figure:appendix2d} in the Appendix, peripheral observations are more disruptive (represented with DI1nok), corresponding to the location of our smaller teams in Figure \ref{figure:team_3d} right panel. Teams consolidating science, as indicated by the Depth variable, are also, on average, the most prominent teams. Small teams that disrupt science tend to have exploratory profiles and/or diverse team compositions. Science disruption seems to occur through small teams with either highly distinct skills or very exploratory profiles. What seems essential is the ability to access a broader knowledge space, regardless of whether this space is reached through the team's highly exploratory profiles or the team's diversity. Teams composed of individuals who are, on average, highly exploratory but with low team cognitive diversity represent teams with similar skills that cover the knowledge space effectively. In contrast, highly diverse teams with specialized individuals also span the knowledge space to propose disruptive ideas, although they may face communication challenges. The combination of these two factors also appears to contribute to disruption, albeit less prominently, suggesting the detrimental effect of excessive diversity.
Another inverted U-shaped relationship exists between a team's average exploratory profile and novelty indicators. When balanced by a relatively exploratory average profile, cognitive diversity appears beneficial without showing a saturation point.

\begin{figure}[h!]
  \centering
  \includegraphics[width=1.05\textwidth]{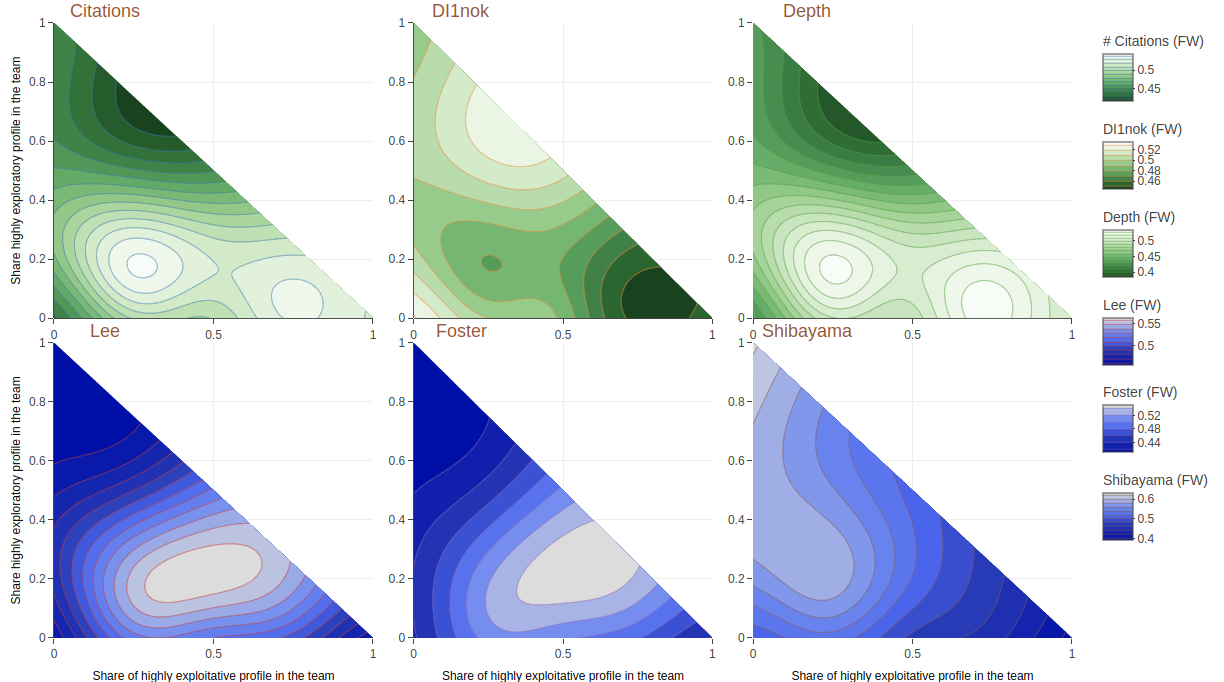}
  \caption{Relation between the share of highly exploitative and highly exploratory profile in a team with and Novelty/ Scientific Impact}
  \label{figure:typic_explo}
\end{figure}

The relationship differs when we adopt an alternative perspective and consider the proportion of highly exploratory and exploitative individuals within scientific teams. A dome is visible in each indicator, signifying successful trade-offs between exploitation and exploration. Figure \ref{figure:typic_explo} offers insight into the relationship between these two aspects and scientific recognition and combinatorial novelty. Teams with fewer highly exploratory individuals and a higher proportion of highly exploitative individuals typically contribute to consolidating the field (Depth metric). Conversely, groups with a higher proportion of highly exploratory individuals and a smaller proportion of highly exploitative individuals are more likely to initiate disruptions in their fields (DI1nok metric). These observations complement the findings of \cite{uzzi2013atypical}, which suggest that a balance between conventional and atypical knowledge combinations produces the most impactful research. Moreover, this analysis enables us to examine how the balance between exploratory and exploitative individuals affects knowledge creation itself. 

Teams featuring a fair proportion of exploratory individuals and a more sustained level of exploitative individuals seem to be most likely to generate compelling new combinations of knowledge. Figure \ref{figure:typic_explo} suggests that an optimal team composition would consist of approximately 50\% highly exploitative and 20\% highly exploratory individuals to increase the likelihood of combining distant knowledge. The situation is less clear for \cite{shibayama2021measuring}, where a high proportion of highly exploratory individuals appears to be beneficial\footnote{\justifying This might be connected with the relationship between our measure and the measure of \cite{shibayama2021measuring} as it is measured in a similar manner. Self-citation also directly impacts the relationship between these two metrics as the same combination of articles will be calculated in both metrics.}. Exploratory individuals contribute to the team by introducing fresh and innovative ideas from their extensive knowledge. These individuals can challenge conventional thinking and steer the team in new directions. Simultaneously, they might foster communication among group members with distant knowledge. In contrast, highly exploitative individuals are crucial for refining and optimizing these novel ideas. Their specialized expertise allows the team to identify feasible and effective solutions, ensuring the creative potential of the exploratory individuals is appropriately channelled into tangible outputs. Additionally, their deep understanding of a specific field facilitates effective communication. The highly exploratory profile complements the specialized knowledge and proficiency of the highly exploitative team members. This dynamic enables the team to capitalize on the full potential of their diverse cognitive abilities, optimizing the innovation process and yielding scientific advancements.

\section{Results}
\label{results}

\subsection{Cognitive dimension and novelty} 
\textbf{Realized novelty}

This subsection examines the relationship between the team's cognitive dimension and novelty indicators. To this end, we report the results of an OLS to identify the joint impact of authors' intra-diversity and inter-diversity on the indicators. The outcomes of these models are presented in Table \ref{ref_cog_nov_fw}. 

\begin{table}[h!]\footnotesize
\centering
\renewcommand{\arraystretch}{1}
\setlength{\tabcolsep}{0.3pt} 
  \caption{Combinatorial Novelty: cognitive diversity and average exploratory profile (Field-Weighted/ References)} 
  \label{} 
\scalebox{1.05}{
	\begin{threeparttable}
\begin{tabular}{@{\extracolsep{5pt}}lccccc} 
\\[-1.8ex]\hline 
\hline \\[-1.8ex] 
 & \multicolumn{5}{c}{\textit{Dependent variable:}} \\ 
\cline{2-6} 
\\[-1.8ex] & Uzzi & Lee & Foster & Wang & Shibayama\\ 
\\[-1.8ex] & \multicolumn{1}{c}{(1)} & \multicolumn{1}{c}{(2)} & \multicolumn{1}{c}{(3)} & \multicolumn{1}{c}{(4)} & \multicolumn{1}{c}{(5)}\\ 
\hline \\[-1.8ex] 
 Author inter $_{abs}$ (FW) & 0.169$^{***}$ & 0.166$^{***}$ & 0.116$^{***}$ & 0.098$^{***}$ & 0.284$^{***}$ \\ 
  & (0.008) & (0.007) & (0.010) & (0.006) & (0.007) \\ 
  & & & & & \\ 
 Author inter $_{abs}\hat{\mkern6mu}$2  (FW) & -0.031$^{***}$ & -0.034$^{***}$ & -0.023$^{**}$ & -0.028$^{***}$ & -0.118$^{***}$ \\ 
  & (0.007) & (0.007) & (0.009) & (0.006) & (0.007) \\ 
  & & & & & \\ 
 Author intra $_{abs}$ (FW) & 0.056$^{***}$ & 0.043$^{***}$ & 0.041$^{**}$ & -0.002 & 0.188$^{***}$ \\ 
  & (0.014) & (0.013) & (0.019) & (0.008) & (0.009) \\ 
  & & & & & \\ 
 Author intra $_{abs}\hat{\mkern6mu}$2  (FW) & -0.088$^{***}$ & -0.094$^{***}$ & -0.084$^{***}$ & -0.026$^{***}$ & -0.047$^{***}$ \\ 
  & (0.011) & (0.010) & (0.015) & (0.006) & (0.010) \\ 
  & & & & & \\ 
 \# References & 0.002$^{***}$ & 0.002$^{***}$ & 0.001$^{***}$ & 0.005$^{***}$ & 0.002$^{***}$ \\ 
  & (0.0001) & (0.0001) & (0.0001) & (0.0001) & (0.0001) \\ 
  & & & & & \\ 
 \# Meshterms & 0.004$^{***}$ & 0.006$^{***}$ & 0.005$^{***}$ & -0.001$^{***}$ & 0.004$^{***}$ \\ 
  & (0.0004) & (0.0004) & (0.0004) & (0.0002) & (0.0004) \\ 
  & & & & & \\ 
 \# Authors & 0.008$^{***}$ & 0.007$^{***}$ & 0.007$^{***}$ & 0.001$^{***}$ & 0.007$^{***}$ \\ 
  & (0.0004) & (0.0004) & (0.0005) & (0.0003) & (0.0003) \\ 
  & & & & & \\ 
 SJR & -0.012$^{***}$ & -0.011$^{***}$ & -0.014$^{***}$ & -0.008$^{***}$ & -0.011$^{***}$ \\ 
  & (0.002) & (0.002) & (0.002) & (0.001) & (0.001) \\ 
  & & & & & \\ 
  Year & Yes & Yes & Yes & Yes & Yes  \\ 
  & & & & & \\ 
  Journal Cat. & Yes & Yes & Yes & Yes & Yes \\ 
  & & & & &  \\
\hline \\[-1.8ex] 
Observations & \multicolumn{1}{c}{1,647,430} & \multicolumn{1}{c}{1,815,603} & \multicolumn{1}{c}{1,815,603} & \multicolumn{1}{c}{1,815,603} & \multicolumn{1}{c}{1,809,155} \\ 
R$^{2}$ & \multicolumn{1}{c}{0.055} & \multicolumn{1}{c}{0.062} & \multicolumn{1}{c}{0.039} & \multicolumn{1}{c}{0.122} & \multicolumn{1}{c}{0.130} \\ 
Adjusted R$^{2}$ & \multicolumn{1}{c}{0.055} & \multicolumn{1}{c}{0.062} & \multicolumn{1}{c}{0.039} & \multicolumn{1}{c}{0.122} & \multicolumn{1}{c}{0.130} \\ 
Residual Std. Error & \multicolumn{1}{c}{0.281 } & \multicolumn{1}{c}{0.278} & \multicolumn{1}{c}{0.310} & \multicolumn{1}{c}{0.345} & \multicolumn{1}{c}{0.267 } \\ 
F Statistic & \multicolumn{1}{c}{406.544$^{***}$ } & \multicolumn{1}{c}{512.283$^{***}$ } & \multicolumn{1}{c}{315.079$^{***}$ } & \multicolumn{1}{c}{1,065.575$^{***}$ } & \multicolumn{1}{c}{1,143.840$^{***}$ } \\ 
\hline 
\hline \\[-1.8ex] 

\end{tabular} 
\begin{tablenotes}
 \footnotesize
 \justifying \item {\it Notes:}
 This table reports coefficients of the effect of cognitive diversity and average exploratory profile on combinatorial novelty using PKG. Standard errors are cluster robust at the journal level: ***, ** and * indicate significance at the 1\%, 5\% and 10\% levels, respectively. The effects are estimated with an OLS. Variables are field-weighted and constant term, scientific field (Scimago Journal Category), and time-fixed effects are incorporated in all model specifications.
 \end{tablenotes}
 \end{threeparttable}
 }
\label{ref_cog_nov_fw}
\end{table} 

First, we confirm that cognitive diversity in a scientific team fosters realized novelty.
Team cognitive diversity (Row 1-2) reveals a significant positive effect on combinatorial novelty. This suggests distant individuals can ease the combination of distant journals in the references. The squared term has negative coefficients. However, the turning point is higher than 1, meaning the relationship is strictly increasing (See Table \ref{turning_points_imp} in Appendix). However, it means that the marginal benefit of cognitive distance is decreasing. 
When interpreting the coefficients, it is important to remember that the independent and dependent variables are expressed in percentile rank within a given field and year. A one percentage point increase in the independent variable's percentile rank implies a $\beta$ percentage point increase in the dependent variable. In our case, the marginal effect of a quadratic term depends on the value of the independent variable. We can calculate marginal effects at the mean values of the independent variable. For example, in \cite{uzzi2013atypical} (model 1), the marginal effect of Author inter$_{abs}$ (FW) at the mean value is calculated this way: $\frac{\Delta y}{\Delta(inter)} = 0.169 - 2 * (-0.031) * Mean(Inter)$. Since variables are expressed in percentile rank, the mean and the median are 0.5. The marginal effect can be then calculated easily, $\frac{\Delta y}{\Delta(inter)} = 0.169 - * (-0.031) = 0.2$. This means that by increasing one percentage point on the ranking of team diversity in a given field and year, one can increase by 0.2 percentage points in the ranking of the most novel articles in the field and year.

On the contrary, the average exploratory profile must remain reasonable to maximize novelty. As visible in Table \ref{turning_points_imp}, the turning points are around 30\% for all indicators except \cite{shibayama2021measuring}, for which it is upper than one. This can mean two things, and this is what we will examine in the second part of this results section, either the researchers have a rather moderate exploratory profile, or there is a balance between exploratory and exploitative individuals. A set of profiles that are too exploratory seems detrimental, as does a set of too exploitative profiles. As shown in Table \ref{ref_cog_nov_fw}, this holds for all indicators on references, except for \cite{wang2017bias}, for which the individual effect is negative, one explanation can be the fact that \cite{wang2017bias} control for future reutilization of the novel combination. Indeed this gives a 'scientific impact' dimension to the metrics and the presence of more specialized individuals may impact the relevance of the combination for the community, making it more likely to be reused.

On MeSH terms, as visible in Table \ref{mesh_cog_nov_fw_pkg} in the Appendix, individual exploratory aspects appear to have a direct negative impact. Indexers assign the MeSH terms and may be subject to bias or misinterpretation. In contrast, the references directly relate to the researchers' choices and reflect their interests and preferences. There are two possibilities, indexers may be unable to capture all the nuances and subtleties of research conducted by individuals with high-average exploratory profiles. Alternatively, the novelty of references could be induced by an author bias in citing previous works irrelevant to the contribution. Researchers' past publications do not directly impact indexers, so she might not need to qualify the article with distant MeSH terms because the novelty is not sufficiently explicit. This suggests that MeSH terms do not reflect the diversity of knowledge and ideas present in individual past work but rather the diversity of competencies between team members.

These relations remain consistent when regressions are not performed using percentage rank information, and indicator behavior with MeSH terms and references seems to be much more corroborated, as visible in Table \ref{ref_cog_nov_pkg} and \ref{mesh_cog_nov_pkg} in the Appendix. The fact that the effect is nearly the same on most of the indicators of novelty demonstrates the robustness of this analysis - our measure captures something similar regardless of the construction of the novelty indicator and the information used.\\

The potential for novelty seems more apparent when looking at the exact composition in terms of exploratory profiles, i.e., the share of exploratory individuals and the share of highly exploitative individuals.
In Table \ref{ref_share_nov_fw_pkg}, we replace the average exploratory profile variables with the exploitative and exploratory individual shares and the interaction of these two variables.

\begin{table}[h!]\footnotesize \centering 
\renewcommand{\arraystretch}{0.8}
\setlength{\tabcolsep}{0.3pt}
  \caption{Combinatorial Novelty: Cognitive diversity, highly exploratory and exploitative profile (Field-Weighted/ References)} 
\label{ref_share_nov_fw_pkg}
    \scalebox{1}{
	\begin{threeparttable}
\begin{tabular}{@{\extracolsep{5pt}}lccccc} 
\\[-1.8ex]\hline 
\hline \\[-1.8ex] 
 & \multicolumn{5}{c}{\textit{Dependent variable:}} \\ 
\cline{2-6} 
\\[-1.8ex] & Uzzi & Lee & Foster & Wang & Shibayama \\ 
\\[-1.8ex] & \multicolumn{1}{c}{(1)} & \multicolumn{1}{c}{(2)} & \multicolumn{1}{c}{(3)} & \multicolumn{1}{c}{(4)} & \multicolumn{1}{c}{(5)}\\ 
\hline \\[-1.8ex] 
 Author inter $_{abs}$ (FW) & 0.168$^{***}$ & 0.163$^{***}$ & 0.107$^{***}$ & 0.066$^{***}$ & 0.400$^{***}$ \\ 
  & (0.014) & (0.012) & (0.020) & (0.008) & (0.012) \\ 
  & & & & & \\ 
  Author inter $_{abs}\hat{\mkern6mu}$2  (FW) & -0.007 & -0.006 & 0.028 & 0.001 & -0.160$^{***}$ \\ 
  & (0.012) & (0.011) & (0.018) & (0.007) & (0.012) \\ 
  & & & & & \\ 
 Share exploratory & -0.166$^{***}$ & -0.173$^{***}$ & -0.214$^{***}$ & -0.084$^{***}$ & -0.022$^{***}$ \\ 
  & (0.007) & (0.007) & (0.010) & (0.004) & (0.006) \\ 
  & & & & & \\ 
 
 Share exploitative & 0.027$^{***}$ & 0.053$^{***}$ & 0.057$^{***}$ & 0.002 & -0.092$^{***}$ \\ 
  & (0.003) & (0.003) & (0.005) & (0.002) & (0.004) \\ 
  & & & & & \\ 
 Share exploratory * Share exploitative & 0.298$^{***}$ & 0.273$^{***}$ & 0.390$^{***}$ & 0.080$^{***}$ & -0.112$^{***}$ \\ 
  & (0.016) & (0.016) & (0.020) & (0.011) & (0.018) \\ 
  & & & & & \\ 
 \# References & 0.002$^{***}$ & 0.002$^{***}$ & 0.001$^{***}$ & 0.005$^{***}$ & 0.002$^{***}$ \\ 
  & (0.0001) & (0.0001) & (0.0001) & (0.0001) & (0.0001) \\ 
  & & & & & \\ 
 \# Meshterms & 0.004$^{***}$ & 0.006$^{***}$ & 0.005$^{***}$ & -0.001$^{***}$ & 0.004$^{***}$ \\ 
  & (0.0004) & (0.0003) & (0.0003) & (0.0002) & (0.0004) \\ 
  & & & & & \\ 
 \# Authors & 0.008$^{***}$ & 0.007$^{***}$ & 0.007$^{***}$ & 0.001$^{***}$ & 0.006$^{***}$ \\ 
  & (0.0004) & (0.0004) & (0.0005) & (0.0002) & (0.0003) \\ 
  & & & & & \\ 
 SJR & -0.012$^{***}$ & -0.011$^{***}$ & -0.014$^{***}$ & -0.008$^{***}$ & -0.011$^{***}$ \\ 
  & (0.002) & (0.001) & (0.002) & (0.001) & (0.001) \\ 
  & & & & & \\ 
  Year & Yes & Yes & Yes & Yes & Yes  \\ 
  & & & & & \\ 
  Journal Cat. & Yes & Yes & Yes & Yes & Yes \\ 
  & & & & &  \\
\hline \\[-1.8ex] 
Observations & \multicolumn{1}{c}{1,647,430} & \multicolumn{1}{c}{1,815,603} & \multicolumn{1}{c}{1,815,603} & \multicolumn{1}{c}{1,815,603} & \multicolumn{1}{c}{1,809,155} \\ 
R$^{2}$ & \multicolumn{1}{c}{0.059} & \multicolumn{1}{c}{0.068} & \multicolumn{1}{c}{0.046} & \multicolumn{1}{c}{0.122} & \multicolumn{1}{c}{0.129} \\ 
Adjusted R$^{2}$ & \multicolumn{1}{c}{0.059} & \multicolumn{1}{c}{0.068} & \multicolumn{1}{c}{0.046} & \multicolumn{1}{c}{0.122} & \multicolumn{1}{c}{0.129} \\ 
Residual Std. Error & \multicolumn{1}{c}{0.280 } & \multicolumn{1}{c}{0.277 } & \multicolumn{1}{c}{0.308 } & \multicolumn{1}{c}{0.345 } & \multicolumn{1}{c}{0.267 } \\ 
F Statistic & \multicolumn{1}{c}{436.681$^{***}$ } & \multicolumn{1}{c}{556.617$^{***}$ } & \multicolumn{1}{c}{372.829$^{***}$ } & \multicolumn{1}{c}{1,065.763$^{***}$ } & \multicolumn{1}{c}{1,132.467$^{***}$ } \\ 
\hline 
\hline \\[-1.8ex] 

\end{tabular} 

\begin{tablenotes}
 \footnotesize
 \justifying \item {\it Notes:}
 This table reports coefficients of the effect of cognitive diversity and highly exploratory and exploitative profiles on combinatorial novelty using PKG. Standard errors are cluster robust at the journal level: ***, ** and * indicate significance at the 1\%, 5\% and 10\% levels, respectively. The effects are estimated with an OLS.  Variables are field-weighted and constant term, scientific field (Scimago Journal Category), and time-fixed effects are incorporated in all model specifications.
 \end{tablenotes}
 \end{threeparttable}
 }
\end{table} 

While cognitive diversity appears to be always beneficial to combine new knowledge, the presence of too many exploratory individuals is harmful. Indeed, its presence only becomes beneficial when counterbalanced by a higher share of exploitative individuals. We can clearly see how this trade-off is necessary to create novelty in the regressions. In the same way as before, the coefficients can be interpreted directly, a percentage point increase in the share of highly exploratory individuals increases by $\beta$ percentage point in the ranking of the most novel articles in the field and year.

Exploratory individuals will develop new perspectives that specialized individuals will capitalize on to make them succeed. A larger share of specialized individuals facilitates communication among members if they are in the same field; otherwise, scientists with diverse backgrounds appear to facilitate communication among team members who are cognitively distant \citep{page2007}. This mirrors the "Renaissance" individual of \citep{jones2009burden} or the "transformational leader" of \cite{shin2007educational} who can ease connections between distant members and foster the team's creativity.

Too many such individuals would make the exploration less efficient, and the emerging ideas would potentially not be successfully implemented because the embedding of the conducted research in a scientific paradigm would not be sufficient. The results are similar across novelty indicators, except for \cite{shibayama2021measuring}, in which the best team composition is made from non-exploitative, non-highly exploratory researchers. Table \ref{mesh_share_nov_fw_pkg} in the Appendix shows that the results also hold for indicators based on MeSH terms.

The two sets of results on the impact of cognitive distance and researcher profile show that combining specialized and exploratory profiles is a good proxy for potential novelty as it enhances the realized novelty in the team\footnote{\justifying Table \ref{ref_share_nov_pkg} provided in the Appendix shows that the results are similar when considering un-normalized indicators}. While \cite{uzzi2013atypical} show that this trade-off between conventional and atypical combinations of knowledge is the most impactful, we demonstrate that this idea holds at the team level as well and that these configurations are most likely to achieve atypical combinations.

\textbf{Perceived novelty}

In this subsection, we examine the relationship between the cognitive dimension and novelty as assessed by experts. Specifically, we employ a Logit model to identify the impact of authors' intra-diversity and inter-diversity on the likelihood of being classified in at least one novel category. The results of these models are presented in Table \ref{z_faculty_op_nonew}. The effect of team cognitive diversity plays a positive role in \textit{perceived} novelty, as seen in the first and second specifications. This effect is less clear when considering individual characteristics. The average exploratory profile has a negative impact. In model 3, we can see that our previous results on \textit{realized} novelty (Table \ref{ref_cog_nov_fw}) only holds for the cognitive distance between individuals when tested on \textit{perceived} novelty. In contrast, when examining the specifications with the share of highly exploratory and exploitative individuals, the results corroborate the regressions performed on \textit{realized} novelty. The proportion of highly exploratory individuals has a negative effect. Instead, typical individuals play a positive role, and the intersection of both types of researchers is indeed positive for predicting novelty. Note that in this specification, cognitive diversity between members is no longer significant.

\begin{table}[h!] \centering 
\renewcommand{\arraystretch}{1}
\setlength{\tabcolsep}{0.3pt}
  \caption{Faculty Opinions: cognitive diversity and average exploratory profile, highly exploratory and exploitative profile (Field-Weighted)} 
  \label{z_faculty_op_nonew} 
\small 
    \scalebox{0.95}{
	\begin{threeparttable}
\begin{tabular}{@{\extracolsep{2pt}}lccccc} 
\\[-1.8ex]\hline 
\hline \\[-1.8ex] 
 & \multicolumn{5}{c}{\textit{Dependent variable:}} \\ 
\cline{2-6} 
\\[-1.8ex] & \multicolumn{5}{c}{Novelty Perceived} \\ 
\\[-1.8ex] & \multicolumn{1}{c}{(1)} & \multicolumn{1}{c}{(2)} & \multicolumn{1}{c}{(3)} & \multicolumn{1}{c}{(4)} & \multicolumn{1}{c}{(5)}\\ 
\midrule
Author inter $_{abs}$ (FW) & 0.306$^{**}$ & 0.715$^{*}$ &  &  & 0.330 \\ 
  & (0.126) & (0.388) &  &  & (0.302) \\ 
  & & & & & \\ 
 Author intra $_{abs}$  (FW) & -0.532$^{***}$ & -0.196 &  &  &  \\ 
  & (0.155) & (0.419) &  &  &  \\ 
  & & & & & \\ 
 Author inter $_{abs}\hat{\mkern6mu}$2  (FW) &  & -0.438 &  &  & -0.270 \\ 
  &  & (0.376) &  &  & (0.325) \\ 
  & & & & & \\ 
 Author intra $_{abs}\hat{\mkern6mu}$2  (FW) &  & -0.364 &  &  &  \\ 
  &  & (0.379) &  &  &  \\ 
  & & & & & \\ 
 Share exploratory &  &  & -0.675$^{**}$ & -1.233$^{***}$ & -1.238$^{***}$ \\ 
  &  &  & (0.275) & (0.371) & (0.384) \\ 
  & & & & & \\ 
 Share exploitative &  &  & 0.339$^{***}$ & 0.317$^{***}$ & 0.337$^{***}$ \\ 
  &  &  & (0.117) & (0.118) & (0.115) \\ 
  & & & & & \\ 
 Share exploratory * Share exploitative &  &  &  & 2.360$^{**}$ & 2.289$^{**}$ \\ 
  &  &  &  & (1.052) & (1.062) \\ 
  & & & & & \\ 
 Control variables & YES & YES & YES & YES & YES \\ 
 &  &  &  &  & \\ 
\hline \\[-1.8ex] 
Observations & \multicolumn{1}{c}{12,555} & \multicolumn{1}{c}{12,555} & \multicolumn{1}{c}{12,555} & \multicolumn{1}{c}{12,555} & \multicolumn{1}{c}{12,555} \\ 
Log Likelihood & \multicolumn{1}{c}{-7,076.944} & \multicolumn{1}{c}{-7,073.965} & \multicolumn{1}{c}{-7,072.608} & \multicolumn{1}{c}{-7,070.408} & \multicolumn{1}{c}{-7,069.551} \\ 
AIC & \multicolumn{1}{c}{14,423.890} & \multicolumn{1}{c}{14,421.930} & \multicolumn{1}{c}{14,415.220} & \multicolumn{1}{c}{14,412.820} & \multicolumn{1}{c}{14,415.100} \\ 
\hline 
\hline \\[-1.8ex]  
\end{tabular} 
\begin{tablenotes}
 \footnotesize
 \justifying \item {\it Notes:}
 This table reports coefficients of the effect of cognitive diversity, average exploratory profile, highly exploratory and exploitative profiles on perceived novelty from Faculty Opinions. Standard errors are cluster robust at the journal level in parentheses: ***, ** and * indicate significance at the 1\%, 5\% and 10\% level, respectively. The effects is estimated using a Logit model. Variables are field-weighted and constant term, scientific field (Scimago Journal Category) and time fixed effects are incorporated in all model specifications.
 \end{tablenotes}
 \end{threeparttable}
 }
\end{table} 

However, when examining Table \ref{aut_cog_f1000} in the Appendix, we can see that the effects are quite heterogeneous across labels. We chose the four labels for which more than 1000 papers had been classified to perform the regressions. The effect of cognitive distance between team members is visible in the ``Technical Advance'' category but not significant for the remaining labels. Conversely, in Table \ref{aut_share_f1000}, we can see that the results in terms of exploratory profiles are mainly driven by the 'Interesting Hypothesis' label. Results here are a bit different since we observe a U shape, meaning that highly specialized or highly diverse teams most often publish articles labelled as ``Interesting hypotheses''. Results are quite similar when using Poisson regression and modelling the number of times a paper is labelled in a given category as visible in Table \ref{aut_cog_f1000_poisson} and Table \ref{aut_share_f1000_poisson} in the Appendix.


\subsection{Cognitive dimension and impact}

This subsection examines the relationship between the team's cognitive dimension and impact measures. To this end, we report the results of an OLS to identify the joint impact of authors' intra-diversity and inter-diversity on the indicators. The outcomes of these models are presented in \ref{cog_imp_fw_pkg} and \ref{share_imp_fw_pkg}. 

Our analysis emphasises the need to differentiate the forms of impact to understand better how the cognitive aspect influences scientific recognition. Indeed, we use the traditional indicator of the number of citations and indicators of disruption and consolidation. The composition of the teams has a significant influence on the type of impact of the studies conducted.

The Table \ref{cog_imp_fw_pkg} regression tables indicate a double inverse U-shaped relationship between the cognitive dimension and the number of citations. Table \ref{turning_points_imp} shows that both turning points are around 45\%. Following Uzzi, a too-conventional work might not be as impactful as the contribution is more marginal. Conversely, peers may not sufficiently consider a too-novel study. This phenomenon is reflected in the composition of the teams as we can see in the differences between consolidation and disruption indicators. Indeed, to consolidate, it is necessary to have a team with a low average exploratory profile and low average cognitive distance between members. The relationship is negative for consolidation indicators (DeIn and Depth) for both intra and inter-individual levels; the effect is sometimes captured via quadratic terms. This means that cognitive diversity is negatively related to the fact that papers citing the focal paper also cite each other or cite many of the references from the focal article. Specialized teams are the ones who consolidate the science.

\begin{table}[h!]\footnotesize \centering 
\renewcommand{\arraystretch}{1}
\setlength{\tabcolsep}{0.1pt}
  \caption{Scientific recognition: cognitive diversity and average exploratory profile (Field-Weighted)} 
  \label{} 
    \scalebox{0.87}{
	\begin{threeparttable}
\begin{tabular}{@{\extracolsep{5pt}}lccccccc} 
\\[-1.8ex]\hline 
\hline \\[-1.8ex] 
 & \multicolumn{7}{c}{\textit{Dependent variable:}} \\ 
\cline{2-8} 
\\[-1.8ex] & \# cit. & DI1 & DI5 & DI1nok & DeIn & Breadth & Depth \\ 
\\[-1.8ex] & \multicolumn{1}{c}{(1)} & \multicolumn{1}{c}{(2)} & \multicolumn{1}{c}{(3)} & \multicolumn{1}{c}{(4)} & \multicolumn{1}{c}{(5)} & \multicolumn{1}{c}{(6)} & \multicolumn{1}{c}{(7)}\\ 
\hline \\[-1.8ex] 
  Author inter $_{abs}$ (FW) & 0.031$^{***}$ & 0.021$^{***}$ & 0.034$^{***}$ & 0.047$^{***}$ & -0.067$^{***}$ & -0.010$^{*}$ & 0.002 \\ 
  & (0.007) & (0.006) & (0.007) & (0.007) & (0.007) & (0.006) & (0.007) \\ 
  & & & & & & & \\
  Author inter $_{abs}\hat{\mkern6mu}$2  (FW) & -0.036$^{***}$ & 0.012$^{**}$ & 0.005 & 0.002 & 0.008 & 0.015$^{***}$ & -0.012$^{**}$ \\ 
  & (0.007) & (0.006) & (0.006) & (0.006) & (0.006) & (0.005) & (0.006) \\ 
  & & & & & & & \\ 
 Author intra $_{abs}$ (FW) & 0.070$^{***}$ & -0.057$^{***}$ & 0.026$^{***}$ & -0.008 & 0.009 & 0.014$^{**}$ & -0.004 \\ 
  & (0.008) & (0.007) & (0.008) & (0.008) & (0.009) & (0.006) & (0.008) \\ 
  & & & & & & & \\ 
 
 Author intra $_{abs}\hat{\mkern6mu}$2  (FW) & -0.072$^{***}$ & 0.038$^{***}$ & 0.009 & 0.024$^{***}$ & -0.030$^{***}$ & 0.021$^{***}$ & -0.038$^{***}$ \\ 
  & (0.008) & (0.007) & (0.007) & (0.007) & (0.007) & (0.006) & (0.007) \\ 
  & & & & & & & \\ 
 \# References & 0.003$^{***}$ & -0.001$^{***}$ & -0.003$^{***}$ & -0.002$^{***}$ & 0.004$^{***}$ & -0.0001$^{*}$ & 0.001$^{***}$ \\ 
  & (0.0001) & (0.0001) & (0.0001) & (0.0001) & (0.0001) & (0.00005) & (0.0001) \\ 
  & & & & & & & \\ 
 \# Meshterms & 0.008$^{***}$ & -0.002$^{***}$ & -0.003$^{***}$ & -0.003$^{***}$ & 0.005$^{***}$ & -0.003$^{***}$ & 0.006$^{***}$ \\ 
  & (0.0004) & (0.0002) & (0.0003) & (0.0003) & (0.0003) & (0.0002) & (0.0004) \\ 
  & & & & & & & \\ 
 \# Authors & 0.012$^{***}$ & -0.006$^{***}$ & -0.002$^{***}$ & -0.005$^{***}$ & 0.006$^{***}$ & -0.009$^{***}$ & 0.012$^{***}$ \\ 
  & (0.0004) & (0.0003) & (0.0003) & (0.0003) & (0.0004) & (0.0003) & (0.0004) \\ 
  & & & & & & & \\ 
 SJR & 0.039$^{***}$ & -0.019$^{***}$ & 0.002 & -0.006$^{***}$ & 0.008$^{***}$ & -0.026$^{***}$ & 0.030$^{***}$ \\ 
  & (0.005) & (0.002) & (0.002) & (0.001) & (0.002) & (0.003) & (0.004) \\ 
  & & & & & & & \\  
  Year & Yes & Yes & Yes & Yes & Yes &  Yes  \\ 
  & & & & & & & \\ 
  Journal Cat. & Yes & Yes & Yes & Yes & Yes &  Yes  \\ 
  & & & & & & & \\ 
\hline \\[-1.8ex] 
Observations & \multicolumn{1}{c}{1,826,207} & \multicolumn{1}{c}{1,826,207} & \multicolumn{1}{c}{1,826,207} & \multicolumn{1}{c}{1,826,207} & \multicolumn{1}{c}{1,826,207} & \multicolumn{1}{c}{1,826,207} & \multicolumn{1}{c}{1,826,207} \\ 
R$^{2}$ & \multicolumn{1}{c}{0.173} & \multicolumn{1}{c}{0.029} & \multicolumn{1}{c}{0.069} & \multicolumn{1}{c}{0.034} & \multicolumn{1}{c}{0.137} & \multicolumn{1}{c}{0.051} & \multicolumn{1}{c}{0.075} \\ 
Adjusted R$^{2}$ & \multicolumn{1}{c}{0.173} & \multicolumn{1}{c}{0.029} & \multicolumn{1}{c}{0.069} & \multicolumn{1}{c}{0.034} & \multicolumn{1}{c}{0.137} & \multicolumn{1}{c}{0.051} & \multicolumn{1}{c}{0.075} \\ 
Residual Std. Error & \multicolumn{1}{c}{0.266} & \multicolumn{1}{c}{0.281} & \multicolumn{1}{c}{0.281} & \multicolumn{1}{c}{0.280} & \multicolumn{1}{c}{0.270} & \multicolumn{1}{c}{0.270} & \multicolumn{1}{c}{0.291} \\ 
F Statistic & \multicolumn{1}{c}{1,621.946$^{***}$} & \multicolumn{1}{c}{233.770$^{***}$} & \multicolumn{1}{c}{575.115$^{***}$} & \multicolumn{1}{c}{269.625$^{***}$} & \multicolumn{1}{c}{1,227.699$^{***}$} & \multicolumn{1}{c}{413.321$^{***}$} & \multicolumn{1}{c}{629.396$^{***}$} \\ 
\hline 
\hline \\[-1.8ex] 
 
\end{tabular} 
\begin{tablenotes}
 \footnotesize
 \justifying \item {\it Notes:}
 This table reports coefficients of the effect of cognitive diversity and average exploratory profile on scientific recognition using PKG. Standard errors are cluster robust at the journal level: ***, ** and * indicate significance at the 1\%, 5\% and 10\% levels, respectively. The effects are estimated with an OLS. Variables are field-weighted and constant term, scientific field (Scimago Journal Category), and time-fixed effects are incorporated in all model specifications.
 \end{tablenotes}
 \end{threeparttable}
 }
\label{cog_imp_fw_pkg}

\end{table} 

For disruptive indicators, the picture is rather different (DI1, DI5, DI1nok and Breadth). Cognitive distance still seems to be globally favorable for disruption. Then, the Breadth disruption indicator, which examines how often articles citing the focal paper also cite each other, seems to indicate a U-shaped relationship with a turning point at 0.33, i.e. if the individuals are very distant or if they are very close, this produces the most disruptive articles in the sense that the citations will be concentrated towards the focal paper.

Although not always significant, the intra-individual effect is more mixed; teams with higher average exploratory profiles globally appear to have a higher disruption potential, but this does not hold for DI1. The DI1NOK index follows the same pattern as DI5, with the exception that it is the quadratic term that takes over.

The articles that are consolidating science are articles with low team diversity and low average exploratory profiles. Here we can observe the notion of highly specialized individuals who conduct more confirmatory and therefore consolidating research. The opposite is true for disruption. The teams' diversity always seems beneficial for proposing disruptive ideas. Articles receiving the most citation are again a matter of a trade-off between a cognitively not-too-distant team and a somewhat reasonable average level of exploration\footnote{\justifying  For regressions without field-year normalization as presented in Table \ref{cog_imp_pkg}, the results are more mixed and less clear. The cognitive aspect seems to follow a U-shaped pattern, with teams that are very close or distant being the most disruptive. The results are more robust for breadth, with diversity consistently appearing to be beneficial.}.

In Table \ref{share_imp_fw_pkg}, we specify the team's composition in terms of exploratory/exploitative profile and found that the relationship of the cognitive distance with the impact measures remains almost similar. For consolidation metrics and citation counts, the share of exploitative individuals is clearly beneficial. The exploitative profile reduces the risk of failure as researchers learn from experience and combinations that have failed \citep{vincenti1990engineers}.
Whereas too exploratory profiles seem to affect the expected number of citations negatively, the effect appears mixed for consolidation since it is positive DeIn and insignificant for Depth. In both cases, combining the two types of profiles is harmful. At the same time, the share of exploitative individuals is positive, suggesting that combining these two types of profiles is not optimal for consolidating research.
To achieve disruption, it is better to minimize the number of individuals who are too exploratory or too specialized, but combining both types of profiles seems once again essential. We can see how the impact of highly exploratory profiles is always negative, and the impact of exploitative profiles is also negative. Still, the interaction between the two is always positive for all disruptiveness measures.

\begin{table}[h!]\footnotesize \centering 

\renewcommand{\arraystretch}{1}
\setlength{\tabcolsep}{0.1pt}
  \caption{Scientific recognition: cognitive diversity, highly exploratory and exploitative profile (Field-Weighted)} 
  \label{} 
    \scalebox{0.78}{
	\begin{threeparttable}
\begin{tabular}{@{\extracolsep{5pt}}lccccccc} 
\\[-1.8ex]\hline 
\hline \\[-1.8ex] 
 & \multicolumn{7}{c}{\textit{Dependent variable:}} \\ 
\cline{2-8} 
\\[-1.8ex] &  \# cit. & DI1 & DI5 & DI1nok & DeIn & Breadth & Depth \\ 
\\[-1.8ex] & \multicolumn{1}{c}{(1)} & \multicolumn{1}{c}{(2)} & \multicolumn{1}{c}{(3)} & \multicolumn{1}{c}{(4)} & \multicolumn{1}{c}{(5)} & \multicolumn{1}{c}{(6)} & \multicolumn{1}{c}{(7)}\\ 
\hline \\[-1.8ex] 
 Author inter $_{abs}$ (FW) & 0.088$^{***}$ & -0.058$^{***}$ & 0.020$^{*}$ & 0.004 & -0.019 & -0.004 & 0.003 \\ 
  & (0.010) & (0.009) & (0.011) & (0.010) & (0.012) & (0.007) & (0.009) \\ 
  & & & & & & & \\ 
  Author inter $_{abs}\hat{\mkern6mu}$2  (FW) & -0.073$^{***}$ & 0.067$^{***}$ & 0.026$^{***}$ & 0.042$^{***}$ & -0.037$^{***}$ & 0.025$^{***}$ & -0.028$^{***}$ \\ 
  & (0.010) & (0.009) & (0.009) & (0.008) & (0.009) & (0.007) & (0.008) \\ 
  & & & & & & & \\ 
 Share exploratory & -0.023$^{***}$ & -0.055$^{***}$ & -0.041$^{***}$ & -0.056$^{***}$ & 0.058$^{***}$ & -0.006 & -0.003 \\ 
  & (0.006) & (0.005) & (0.006) & (0.005) & (0.006) & (0.004) & (0.005) \\ 
  & & & & & & & \\ 
 
 Share exploitative & 0.029$^{***}$ & -0.033$^{***}$ & -0.056$^{***}$ & -0.049$^{***}$ & 0.058$^{***}$ & -0.024$^{***}$ & 0.032$^{***}$ \\ 
  & (0.003) & (0.003) & (0.003) & (0.002) & (0.003) & (0.002) & (0.003) \\ 
  & & & & & & & \\ 
 Share exploratory * Share exploitative & -0.023$^{**}$ & 0.132$^{***}$ & 0.047$^{***}$ & 0.096$^{***}$ & -0.087$^{***}$ & 0.059$^{***}$ & -0.034$^{***}$ \\ 
  & (0.012) & (0.010) & (0.011) & (0.010) & (0.011) & (0.011) & (0.012) \\ 
  & & & & & & & \\ 
 \# References & 0.003$^{***}$ & -0.001$^{***}$ & -0.003$^{***}$ & -0.002$^{***}$ & 0.004$^{***}$ & -0.0001$^{*}$ & 0.001$^{***}$ \\ 
  & (0.0001) & (0.0001) & (0.0001) & (0.0001) & (0.0001) & (0.00005) & (0.0001) \\ 
  & & & & & & & \\ 
 \# Meshterms & 0.008$^{***}$ & -0.002$^{***}$ & -0.003$^{***}$ & -0.003$^{***}$ & 0.005$^{***}$ & -0.003$^{***}$ & 0.006$^{***}$ \\ 
  & (0.0004) & (0.0002) & (0.0003) & (0.0003) & (0.0003) & (0.0002) & (0.0004) \\ 
  & & & & & & & \\ 
 \# Authors & 0.012$^{***}$ & -0.007$^{***}$ & -0.003$^{***}$ & -0.005$^{***}$ & 0.006$^{***}$ & -0.010$^{***}$ & 0.013$^{***}$ \\ 
  & (0.0004) & (0.0003) & (0.0003) & (0.0003) & (0.0004) & (0.0003) & (0.0004) \\ 
  & & & & & & & \\ 
 SJR & 0.038$^{***}$ & -0.018$^{***}$ & 0.003 & -0.006$^{***}$ & 0.007$^{***}$ & -0.026$^{***}$ & 0.030$^{***}$ \\ 
  & (0.005) & (0.002) & (0.002) & (0.001) & (0.002) & (0.003) & (0.004) \\ 
  & & & & & & & \\ 
  Year & Yes & Yes & Yes & Yes & Yes &  Yes & Yes \\ 
  & & & & & & & \\ 
  Journal Cat. & Yes & Yes & Yes & Yes & Yes & Yes & Yes  \\ 
  & & & & & & & \\ 
\hline \\[-1.8ex] 
Observations & \multicolumn{1}{c}{1,826,207} & \multicolumn{1}{c}{1,826,207} & \multicolumn{1}{c}{1,826,207} & \multicolumn{1}{c}{1,826,207} & \multicolumn{1}{c}{1,826,207} & \multicolumn{1}{c}{1,826,207} & \multicolumn{1}{c}{1,826,207} \\ 
R$^{2}$ & \multicolumn{1}{c}{0.174} & \multicolumn{1}{c}{0.030} & \multicolumn{1}{c}{0.071} & \multicolumn{1}{c}{0.035} & \multicolumn{1}{c}{0.139} & \multicolumn{1}{c}{0.051} & \multicolumn{1}{c}{0.075} \\ 
Adjusted R$^{2}$ & \multicolumn{1}{c}{0.174} & \multicolumn{1}{c}{0.030} & \multicolumn{1}{c}{0.071} & \multicolumn{1}{c}{0.035} & \multicolumn{1}{c}{0.139} & \multicolumn{1}{c}{0.050} & \multicolumn{1}{c}{0.075} \\ 
Residual Std. Error & \multicolumn{1}{c}{0.266} & \multicolumn{1}{c}{0.281} & \multicolumn{1}{c}{0.280} & \multicolumn{1}{c}{0.280} & \multicolumn{1}{c}{0.270} & \multicolumn{1}{c}{0.270} & \multicolumn{1}{c}{0.291} \\ 
F Statistic  & \multicolumn{1}{c}{1,619.636$^{***}$} & \multicolumn{1}{c}{239.244$^{***}$} & \multicolumn{1}{c}{586.510$^{***}$} & \multicolumn{1}{c}{281.922$^{***}$} & \multicolumn{1}{c}{1,243.711$^{***}$} & \multicolumn{1}{c}{410.608$^{***}$} & \multicolumn{1}{c}{626.296$^{***}$} \\ 
\hline 
\hline \\[-1.8ex] 

\end{tabular} 
\begin{tablenotes}
 \footnotesize
 \justifying \item {\it Notes:}
 This table reports coefficients of the effect of cognitive diversity and highly exploratory and exploitative profiles on scientific recognition using PKG. Standard errors are cluster robust at the journal level: ***, ** and * indicate significance at the 1\%, 5\% and 10\% levels, respectively. The effects are estimated with an OLS.  Variables are field-weighted and constant term, scientific field (Scimago Journal Category), and time-fixed effects are incorporated in all model specifications.
 \end{tablenotes}
 \end{threeparttable}
 }
 \label{share_imp_fw_pkg}
\end{table} 

In conclusion, the analysis shows how teams with a high share of specialized individuals or low average exploratory profiles are teams that consolidate science. In contrast, teams that get the most recognition in terms of disruption combine highly exploitative and highly exploratory individuals and have cognitively more distant members\footnote{\justifying   For regressions without field-year normalization (see Table \ref{share_imp_pkg}), the results are less homogeneous for the cognitive distance aspect, but the combination of exploratory and exploitative is robust. The interaction of the two consistently leads to disruption.}.

\section{Conclusion}
\label{ccl}

This paper examines the effect of exploratory scholars and, in a broader way, team composition on creativity. Our findings suggest that the cognitive dimension plays a crucial role in the creative process, and significantly influences the two pillars of creativity: originality and success. We first show that the team's cognitive diversity strongly influences novelty (\textit{realized} and \textit{perceived}) of the research conducted. We also show that a double-inversed U-shaped relationship exists between cognitive dimensions (intra and inter) and the impact in terms of citations. Our study also highlights the strong connection between the cognitive dimension and the nature of these citations. Teams with more exploitative profiles tend to consolidate science, while those with more exploratory individuals disrupt it and propose more distant knowledge combinations, only when associated with exploitative ones. Our research underscores how team composition in terms of profiles lies at the heart of scientific creativity.

Multiple limitations arise in our study. First, concerning data used, PKG is based on advanced heuristics and algorithms to disambiguate authors using affiliation and additional metadata \cite{xu2020building}. While there is a considerable amount of research on addressing noise in Knowledge Graphs \citep{fasoulis2020error} and improvements in these methods may increase their reliability in the future, we cannot guarantee that errors or inconsistencies will not occur when dealing with author-level information in PKG. 

Other shortcomings are directly related to the creation of our indicator. First, many methods and hyper-parameters were chosen for the simplicity of computation. The embedding is a pre-trained model from SpaCy and is not state-of-the-art. One should compare the behavior of different embedding techniques but also on what kind of text they are applied and the distance measure used. We suspect that the two papers might be close given a specific embedding and distance measure but highly distant given other parameters. In addition, the distance between the two papers would vary depending on whether the distance metric is applied to the paper's title, abstract, or full text. The semantic distances between researchers can be influenced by biases inherent in the fields and journal practices. For example, if researchers publish in different journals, the structure and format of their abstracts may be affected even if their research topic or area of expertise remains unchanged. Another hyper-parameter we used is the time window for an author's past publication. We considered a time window of 5 years. This suggests that any paper published by the author before this point would not be captured. One could argue that past behavior influences current behavior, and a highly diverse background can be proxied by recent publications. Yet no evidence supports this hypothesis. Another issue is how we define authors' cognitive aspect by considering only past publications. Although we do not try to approximate the skills of a researcher but only their disposition to do diverse research, we are not sure how working on a topic is enough to understand then and manage this new knowledge. This raises the question of the exact competencies of a transformational leader and if the past paper is sufficient to proxy it. Also, a specialized author could have previously worked on distant papers but only on his topic/methodology. Our measure defines it as diverse, yet is it true? Although solo publications can be used to construct an author's profile, the increasing significance of teamwork in scientific research makes it uncertain whether a complete and precise profile can be established solely on this basis. Another option could be to incorporate external information, such as educational background, and assign greater weight to papers that align with the author's education. However, obtaining this information can be challenging as it often requires web scraping, which is not easily scalable. The last issue in our mind about using past publications is ghost and honorary authorship as it is common that some authors contributed very little to the production of the article. \citep{sugimoto2018measuring,pruschak2022and}. Both are problems to consider while defining a coauthored paper as part of your knowledge space.

In our analysis, we solely focused on the cognitive diversity of researchers, but diversity encompasses various aspects as highlighted by prior research studies \citep{medin2012diversity, hofstra2020diversity}. According to \cite{koopmann2021proximity},  there are four proximity dimensions among researchers, namely cognitive, institutional, social, and geographical. Relying solely on PKG to approximate all of these dimensions could be challenging. Still, alternative sources such as OpenAlex could provide more comprehensive information on a researcher's institutions, past institutions, and authors' characteristics. For instance, relying on PKG to construct a researcher's seniority could be biased because of the restriction on health sciences papers. Exploring these additional channels could lead to developing supplementary measures that complement cognitive diversity.

Another area worth exploring is the temporal dynamic between exploring new ideas and exploiting existing ones. As we discussed earlier, discovering new concepts is essential for addressing major challenges. However, there is often a pattern of moving through cycles of exploration and exploitation within a particular field. Similarly, authors may initially focus on a particular subject and then switch to a different area to gain a fresh perspective on the first one once they have developed sufficient expertise.
 
To increase the efficiency of the scientific system, it is necessary to conduct further research on the composition of research teams and their impact on creativity. Our preliminary results indicate that policymakers and grant evaluators should consider both individual and team-level characteristics and not only success or originality when making decisions about research funding and support. We have explored some research avenues to deepen our understanding of this phenomenon, and we encourage other researchers to build upon our work in this area. By continuing to investigate these factors, we can develop more effective strategies for supporting and fostering creativity within research teams, ultimately leading to more impactful and innovative scientific outcomes.



\bibliography{ref}

\newpage

\section{Appendix}
\section*{Data and code availability}

The curated dataset used in the analysis described herein can be
accessed through the Zenodo repository https://zenodo.org/records/8382881. The Python and R scripts used to compute indicators, tables and plots reported in the paper are accessible through GitHub at https://github.com/Kwirtz/Unpacking-scientific-creativity.
\section*{Figures}

    \begin{figure}[h!]
      \centering
      \includegraphics[width=1.1\textwidth]{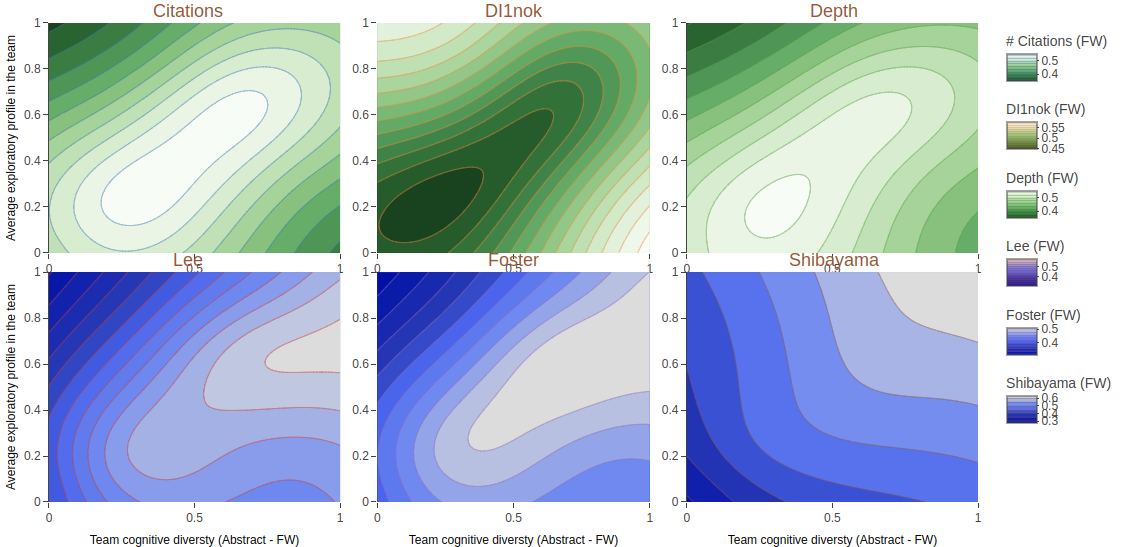}
      \caption{Relation between cognitive diversity, average exploratory profile and Novelty/ Scientific Impact}
      \label{figure:appendix2d}
    \end{figure}

\newpage
\section*{Regressions}

\subsection*{Novelty indicators and Faculty Opinion}

\begin{table}[h!]\footnotesize \centering 
\renewcommand{\arraystretch}{1}
\setlength{\tabcolsep}{0.5pt}
  \caption{Faculty Opinions: Cognitive diversity and average exploratory profile (Field-Weighted)} 
  \label{aut_cog_f1000} 
    \scalebox{1}{
	\begin{threeparttable}
\begin{tabular}{@{\extracolsep{5pt}}lcccc} 
\\[-1.8ex]\hline 
\hline \\[-1.8ex] 
 & \multicolumn{4}{c}{\textit{Dependent variable:}} \\ 
\cline{2-5} 
\\[-1.8ex] & \multicolumn{4}{c}{Logit Model} \\ 
\\[-1.8ex] & Interesting Hyp. &  Technical Adv. & Confirmation & Controversial \\ 
\hline \\[-1.8ex] 
 Author inter $_{abs}$ (FW) & -0.625 & 1.485$^{***}$ & -0.427 & -0.757 \\ 
  & (0.387) & (0.338) & (0.386) & (0.491) \\ 
  & & & & \\ 
   Author inter $_{abs}\hat{\mkern6mu}$2  (FW) & 0.414 & -1.101$^{***}$ & 0.310 & 0.543 \\ 
  & (0.382) & (0.328) & (0.381) & (0.516) \\ 
  & & & & \\ 
 Author intra $_{abs}$ (FW) & -0.191 & 0.209 & 0.001 & 0.278 \\ 
  & (0.388) & (0.336) & (0.384) & (0.580) \\ 
  & & & & \\ 

 Author intra $_{abs}\hat{\mkern6mu}$2  (FW) & -0.016 & -0.465 & 0.199 & 0.016 \\ 
  & (0.383) & (0.324) & (0.365) & (0.602) \\ 
  & & & & \\ 
 \# References & 0.006$^{***}$ & -0.012$^{***}$ & -0.0002 & -0.001 \\ 
  & (0.001) & (0.002) & (0.001) & (0.002) \\ 
  & & & & \\ 
 \# Meshterms & 0.019$^{***}$ & -0.040$^{***}$ & 0.011$^{***}$ & 0.004 \\ 
  & (0.004) & (0.006) & (0.004) & (0.005) \\ 
  & & & & \\ 
 \# Authors & -0.026$^{***}$ & 0.018$^{***}$ & 0.005 & -0.017$^{*}$ \\ 
  & (0.006) & (0.005) & (0.004) & (0.009) \\ 
  & & & & \\ 
 SJR & 0.065$^{***}$ & -0.019$^{**}$ & -0.008 & 0.006 \\ 
  & (0.011) & (0.010) & (0.005) & (0.008) \\ 
  & & & & \\ 
  Year & Yes & Yes & Yes & Yes  \\ 
  & & & & \\ 
  Journal Cat. & Yes & Yes & Yes & Yes \\ 
  & & & &  \\
\hline \\[-1.8ex] 
Observations & \multicolumn{1}{c}{12,555} & \multicolumn{1}{c}{12,555} & \multicolumn{1}{c}{12,555} & \multicolumn{1}{c}{12,555} \\ 
Log Likelihood & \multicolumn{1}{c}{-7,919.383} & \multicolumn{1}{c}{-7,202.326} & \multicolumn{1}{c}{-7,657.496} & \multicolumn{1}{c}{-3,866.333} \\ 
Akaike Inf. Crit. & \multicolumn{1}{c}{16,112.770} & \multicolumn{1}{c}{14,678.650} & \multicolumn{1}{c}{15,588.990} & \multicolumn{1}{c}{8,006.667} \\ 
\hline 
\hline \\[-1.8ex] 

\end{tabular} 
 \begin{tablenotes}
 \footnotesize
 \justifying \item {\it Notes:}
 This table reports coefficients of the effect of cognitive diversity and average exploratory profile on perceived novelty from Faculty Opinions. Standard errors are cluster robust at the journal-level: ***, ** and * indicate significance at the 1\%, 5\% and 10\% level, respectively. The effects are estimated with a Logit model. Variables are field-weighted and constant term, scientific field (Scimago Journal Category) and time fixed effects are incorporated in all model specifications.
 \end{tablenotes}
 \end{threeparttable}
 }
\end{table} 

\begin{table}[h!] \footnotesize
\centering 
\renewcommand{\arraystretch}{1}
\setlength{\tabcolsep}{0.5pt}
  \caption{Faculty Opinions: Cognitive diversity and average exploratory profile  (Field-Weighted)} 
  \label{aut_cog_f1000_poisson} 
    \scalebox{1}{
	\begin{threeparttable}
\begin{tabular}{@{\extracolsep{5pt}}lcccc} 
\\[-1.8ex]\hline 
\hline \\[-1.8ex] 
 & \multicolumn{4}{c}{\textit{Dependent variable:}} \\ 
\cline{2-5} 
\\[-1.8ex] & \multicolumn{4}{c}{Poisson Model} \\ 
\\[-1.8ex] & Interesting Hyp. &  Technical Adv. & Confirmation & Controversial \\ 
\hline \\[-1.8ex] 
Author inter $_{abs}$ (FW) & -0.410$^{*}$ & 1.225$^{***}$ & -0.330 & -0.828$^{**}$ \\ 
  & (0.209) & (0.204) & (0.287) & (0.413) \\ 
  & & & & \\ 
  Author inter $_{abs}\hat{\mkern6mu}$2  (FW) & 0.291 & -0.918$^{***}$ & 0.259 & 0.689 \\ 
  & (0.210) & (0.192) & (0.285) & (0.440) \\ 
  & & & & \\ 
 Author intra $_{abs}$ (FW) & -0.009 & 0.320 & 0.129 & 0.271 \\ 
  & (0.184) & (0.217) & (0.248) & (0.495) \\ 
  & & & & \\ 
 
 Author intra $_{abs}\hat{\mkern6mu}$2  (FW) & -0.119 & -0.492$^{**}$ & -0.041 & -0.108 \\ 
  & (0.181) & (0.219) & (0.223) & (0.521) \\ 
  & & & & \\ 
 \# References & 0.003$^{***}$ & -0.006$^{***}$ & -0.0001 & 0.0005 \\ 
  & (0.001) & (0.002) & (0.001) & (0.001) \\ 
  & & & & \\ 
 \# Meshterms & 0.013$^{***}$ & -0.024$^{***}$ & 0.007$^{*}$ & 0.002 \\ 
  & (0.002) & (0.005) & (0.003) & (0.005) \\ 
  & & & & \\ 
 \# Authors & -0.017$^{***}$ & 0.009$^{***}$ & 0.003 & -0.012 \\ 
  & (0.004) & (0.003) & (0.003) & (0.009) \\ 
  & & & & \\ 
 SJR & 0.039$^{***}$ & -0.0004 & 0.002 & 0.014$^{*}$ \\ 
  & (0.007) & (0.007) & (0.004) & (0.007) \\ 
  & & & & \\ 
  Year & Yes & Yes & Yes & Yes  \\ 
  & & & & \\ 
  Journal Cat. & Yes & Yes & Yes & Yes \\ 
  & & & &  \\
\hline \\[-1.8ex] 
Observations & \multicolumn{1}{c}{12,555} & \multicolumn{1}{c}{12,555} & \multicolumn{1}{c}{12,555} & \multicolumn{1}{c}{12,555} \\ 
Log Likelihood & \multicolumn{1}{c}{-10,420.250} & \multicolumn{1}{c}{-9,880.221} & \multicolumn{1}{c}{-8,978.963} & \multicolumn{1}{c}{-4,358.803} \\ 
Akaike Inf. Crit. & \multicolumn{1}{c}{21,114.510} & \multicolumn{1}{c}{20,034.440} & \multicolumn{1}{c}{18,231.920} & \multicolumn{1}{c}{8,991.606} \\ 
\hline 
\hline \\[-1.8ex] 

\end{tabular}
 \begin{tablenotes}
 \footnotesize
 \justifying \item {\it Notes:}
 This table reports coefficients of the effect of cognitive diversity and average exploratory profile on perceived novelty from Faculty Opinions. Standard errors are cluster robust at the journal level: ***, ** and * indicate significance at the 1\%, 5\% and 10\% level, respectively. The effects are estimated with a Poisson model. Variables are field-weighted and constant term, scientific field (Scimago Journal Category) and time fixed effects are incorporated in all model specifications.
 \end{tablenotes}
 \end{threeparttable}
 }

\end{table}

\begin{table}[h!]\footnotesize \centering 
\renewcommand{\arraystretch}{1}
\setlength{\tabcolsep}{0.5pt}
  \caption{Faculty Opinions: Cognitive diversity, highly exploratory and exploitative profile (Field-Weighted)} 
  \label{aut_share_f1000} 
    \scalebox{0.95}{
	\begin{threeparttable}
\begin{tabular}{@{\extracolsep{5pt}}lcccc} 
\\[-1.8ex]\hline 
\hline \\[-1.8ex] 
 & \multicolumn{4}{c}{\textit{Dependent variable:}} \\ 
\cline{2-5} 
\\[-1.8ex] & \multicolumn{4}{c}{Logit Model} \\ 
\\[-1.8ex] & Interesting Hyp. &  Technical Adv. & Confirmation & Controversial \\ 
\hline \\[-1.8ex] 
Author inter $_{abs}$ (FW) & -0.859$^{***}$ & 1.500$^{***}$ & -0.289 & -0.476 \\ 
  & (0.277) & (0.317) & (0.309) & (0.377) \\ 
  & & & & \\ 
   Author inter $_{abs}\hat{\mkern6mu}$2  (FW) & 0.590$^{*}$ & -1.244$^{***}$ & 0.228 & 0.421 \\ 
  & (0.308) & (0.338) & (0.333) & (0.371) \\ 
  & & & & \\ 
 Share exploratory & -0.754$^{*}$ & -0.644 & 0.868$^{**}$ & -0.193 \\ 
  & (0.450) & (0.443) & (0.441) & (0.607) \\ 
  & & & & \\ 

 Share exploitative & 0.304$^{***}$ & -0.014 & -0.070 & -0.097 \\ 
  & (0.083) & (0.118) & (0.097) & (0.160) \\ 
  & & & & \\ 
 Share exploratory * Share exploitative & 2.911$^{***}$ & 0.015 & -1.069 & 1.822 \\ 
  & (1.073) & (1.132) & (1.048) & (1.650) \\ 
  & & & & \\ 
 \# References & 0.006$^{***}$ & -0.012$^{***}$ & -0.0001 & -0.001 \\ 
  & (0.001) & (0.002) & (0.001) & (0.002) \\ 
  & & & & \\ 
 \# Meshterms & 0.019$^{***}$ & -0.041$^{***}$ & 0.012$^{***}$ & 0.004 \\ 
  & (0.004) & (0.006) & (0.004) & (0.005) \\ 
  & & & & \\ 
 \# Authors & -0.024$^{***}$ & 0.018$^{***}$ & 0.005 & -0.018$^{*}$ \\ 
  & (0.006) & (0.005) & (0.004) & (0.009) \\ 
  & & & & \\ 
 SJR & 0.065$^{***}$ & -0.019$^{*}$ & -0.008 & 0.005 \\ 
  & (0.010) & (0.010) & (0.005) & (0.008) \\ 
  & & & & \\ 
  Year & Yes & Yes & Yes & Yes  \\ 
  & & & & \\ 
  Journal Cat. & Yes & Yes & Yes & Yes \\ 
  & & & &  \\
\hline \\[-1.8ex] 
Observations & \multicolumn{1}{c}{12,555} & \multicolumn{1}{c}{12,555} & \multicolumn{1}{c}{12,555} & \multicolumn{1}{c}{12,555} \\ 
Log Likelihood & \multicolumn{1}{c}{-7,910.400} & \multicolumn{1}{c}{-7,202.819} & \multicolumn{1}{c}{-7,655.698} & \multicolumn{1}{c}{-3,866.431} \\ 
Akaike Inf. Crit. & \multicolumn{1}{c}{16,096.800} & \multicolumn{1}{c}{14,681.640} & \multicolumn{1}{c}{15,587.400} & \multicolumn{1}{c}{8,008.863} \\ 
\hline 
\hline \\[-1.8ex] 

\end{tabular} 

\begin{tablenotes}
 \footnotesize
 \justifying \item {\it Notes:}
 This table reports coefficients of the effect of cognitive diversity and highly exploratory and exploitative profiles on perceived novelty from Faculty Opinions. Standard errors are cluster robust at the journal-level: ***, ** and * indicate significance at the 1\%, 5\% and 10\% level, respectively. The effects are estimated with a Logit model. Variables are field-weighted and constant term, scientific field (Scimago Journal Category) and time fixed effects are incorporated in all model specifications.
 \end{tablenotes}
 \end{threeparttable}
 }
\end{table} 

\begin{table}[h!]\footnotesize \centering 
\renewcommand{\arraystretch}{1}
\setlength{\tabcolsep}{0.5pt}
  \caption{Faculty Opinions: Cognitive diversity, highly exploratory and exploitative profile (Field-Weighted)} 
  \label{aut_share_f1000_poisson} 
    \scalebox{0.95}{
	\begin{threeparttable}
\begin{tabular}{@{\extracolsep{5pt}}lcccc} 
\\[-1.8ex]\hline 
\hline \\[-1.8ex] 
 & \multicolumn{4}{c}{\textit{Dependent variable:}} \\ 
\cline{2-5} 
\\[-1.8ex] & \multicolumn{4}{c}{Poisson Model} \\ 
\\[-1.8ex] & Interesting Hyp. &  Technical Adv. & Confirmation & Controversial \\ 
\hline \\[-1.8ex] 
 Author inter $_{abs}$ (FW) & -0.485$^{***}$ & 1.336$^{***}$ & -0.171 & -0.560$^{*}$ \\ 
  & (0.157) & (0.224) & (0.248) & (0.322) \\ 
  & & & & \\ 
  Author inter $_{abs}\hat{\mkern6mu}$2  (FW) & 0.318$^{*}$ & -1.106$^{***}$ & 0.121 & 0.472 \\ 
  & (0.163) & (0.224) & (0.261) & (0.332) \\ 
  & & & & \\ 
 Share exploratory & -0.718$^{**}$ & -0.550$^{*}$ & 0.478$^{*}$ & -0.083 \\ 
  & (0.334) & (0.312) & (0.246) & (0.513) \\ 
  & & & & \\ 
 
 Share exploitative & 0.135$^{***}$ & -0.026 & -0.020 & -0.084 \\ 
  & (0.049) & (0.077) & (0.066) & (0.171) \\ 
  & & & & \\ 
 Share exploratory * Share exploitative & 2.112$^{***}$ & -0.106 & -0.564 & 1.674 \\ 
  & (0.662) & (0.762) & (0.640) & (1.504) \\ 
  & & & & \\ 
 \# References & 0.003$^{***}$ & -0.006$^{***}$ & -0.0001 & 0.0005 \\ 
  & (0.001) & (0.002) & (0.001) & (0.001) \\ 
  & & & & \\ 
 \# Meshterms & 0.013$^{***}$ & -0.024$^{***}$ & 0.007$^{**}$ & 0.002 \\ 
  & (0.002) & (0.005) & (0.003) & (0.005) \\ 
  & & & & \\ 
 \# Authors & -0.016$^{***}$ & 0.009$^{***}$ & 0.003 & -0.013 \\ 
  & (0.004) & (0.003) & (0.003) & (0.009) \\ 
  & & & & \\ 
 SJR & 0.039$^{***}$ & 0.00003 & 0.002 & 0.014$^{*}$ \\ 
  & (0.007) & (0.007) & (0.004) & (0.007) \\ 
  & & & & \\ 
  Year & Yes & Yes & Yes & Yes  \\ 
  & & & & \\ 
  Journal Cat. & Yes & Yes & Yes & Yes \\ 
  & & & &  \\
\hline \\[-1.8ex] 
Observations & \multicolumn{1}{c}{12,555} & \multicolumn{1}{c}{12,555} & \multicolumn{1}{c}{12,555} & \multicolumn{1}{c}{12,555} \\ 
Log Likelihood & \multicolumn{1}{c}{-10,415.010} & \multicolumn{1}{c}{-9,880.661} & \multicolumn{1}{c}{-8,977.912} & \multicolumn{1}{c}{-4,358.090} \\ 
Akaike Inf. Crit. & \multicolumn{1}{c}{21,106.010} & \multicolumn{1}{c}{20,037.320} & \multicolumn{1}{c}{18,231.830} & \multicolumn{1}{c}{8,992.180} \\ 
\hline 
\hline \\[-1.8ex] 

\end{tabular} 

\begin{tablenotes}
 \footnotesize
 \justifying \item {\it Notes:}
 This table reports coefficients of the effect of cognitive diversity and highly exploratory and exploitative profiles on perceived novelty from Faculty Opinions. Standard errors are cluster robust at the journal level: ***, ** and * indicate significance at the 1\%, 5\% and 10\% level, respectively. The effects are estimated with a Poisson model. Variables are field-weighted and constant term, scientific field (Scimago Journal Category) and time fixed effects are incorporated in all model specifications.
 \end{tablenotes}
 \end{threeparttable}
 }
\end{table} 

\clearpage

\subsection*{Novelty indicators with Mesh Terms}
\subsubsection*{Cognitive diversity and average exploratory profile effect on Novelty}

\begin{table}[h!]\footnotesize \centering 
\renewcommand{\arraystretch}{1}
\setlength{\tabcolsep}{0.5pt}
  \caption{Combinatorial Novelty: cognitive diversity and average exploratory profile (Field-Weighted/ Meshterms)} 
  \label{} 
    \scalebox{1}{
	\begin{threeparttable}
\begin{tabular}{@{\extracolsep{25pt}}lcccc} 
\\[-1.8ex]\hline 
\hline \\[-1.8ex] 
 & \multicolumn{4}{c}{\textit{Dependent variable:}} \\ 
\cline{2-5} 
\\[-1.8ex] & Uzzi & Lee & Foster & Wang  \\ 
\\[-1.8ex] & \multicolumn{1}{c}{(1)} & \multicolumn{1}{c}{(2)} & \multicolumn{1}{c}{(3)} & \multicolumn{1}{c}{(4)}\\ 
\hline \\[-1.8ex] 
Author inter $_{abs}$ (FW) & 0.067$^{***}$ & 0.114$^{***}$ & 0.056$^{***}$ & 0.062$^{***}$ \\ 
  & (0.009) & (0.007) & (0.008) & (0.006) \\ 
  & & & & \\ 
  Author inter $_{abs}\hat{\mkern6mu}$2  (FW) & -0.016$^{**}$ & -0.050$^{***}$ & -0.025$^{***}$ & -0.008 \\ 
  & (0.008) & (0.006) & (0.008) & (0.006) \\ 
  & & & & \\ 
 Author intra $_{abs}$ (FW) & -0.020$^{*}$ & 0.025$^{***}$ & -0.029$^{**}$ & -0.055$^{***}$ \\ 
  & (0.012) & (0.009) & (0.013) & (0.009) \\ 
  & & & & \\ 
 
 Author intra $_{abs}\hat{\mkern6mu}$2  (FW) & -0.047$^{***}$ & -0.062$^{***}$ & -0.042$^{***}$ & -0.010 \\ 
  & (0.010) & (0.008) & (0.011) & (0.007) \\ 
  & & & & \\ 
 \# References & 0.001$^{***}$ & 0.001$^{***}$ & 0.001$^{***}$ & 0.0002$^{***}$ \\ 
  & (0.0001) & (0.00005) & (0.0001) & (0.00003) \\ 
  & & & & \\ 
 \# Meshterms & 0.007$^{***}$ & 0.014$^{***}$ & 0.0004 & 0.029$^{***}$ \\ 
  & (0.001) & (0.001) & (0.0004) & (0.0004) \\ 
  & & & & \\ 
 \# Authors & 0.002$^{***}$ & 0.008$^{***}$ & 0.005$^{***}$ & 0.001$^{***}$ \\ 
  & (0.0004) & (0.0004) & (0.0004) & (0.0003) \\ 
  & & & & \\ 
 SJR & -0.004$^{***}$ & -0.006$^{***}$ & -0.005$^{***}$ & 0.003$^{***}$ \\ 
  & (0.001) & (0.001) & (0.001) & (0.001) \\ 
  & & & & \\ 
  Year & Yes & Yes & Yes & Yes  \\ 
  & & & &  \\ 
  Journal Cat. & Yes & Yes & Yes & Yes \\ 
  & & & &  \\ 
\hline \\[-1.8ex] 
Observations & \multicolumn{1}{c}{661,821} & \multicolumn{1}{c}{1,823,859} & \multicolumn{1}{c}{1,823,859} & \multicolumn{1}{c}{1,823,859} \\ 
R$^{2}$ & \multicolumn{1}{c}{0.029} & \multicolumn{1}{c}{0.083} & \multicolumn{1}{c}{0.015} & \multicolumn{1}{c}{0.153} \\ 
Adjusted R$^{2}$ & \multicolumn{1}{c}{0.029} & \multicolumn{1}{c}{0.083} & \multicolumn{1}{c}{0.015} & \multicolumn{1}{c}{0.152} \\ 
Residual Std. Error & \multicolumn{1}{c}{0.285 } & \multicolumn{1}{c}{0.276 } & \multicolumn{1}{c}{0.300 } & \multicolumn{1}{c}{0.360 } \\ 
F Statistic & \multicolumn{1}{c}{86.982$^{***}$ } & \multicolumn{1}{c}{699.050$^{***}$ } & \multicolumn{1}{c}{121.183$^{***}$ } & \multicolumn{1}{c}{1,390.452$^{***}$ } \\ 
\hline 
\hline \\[-1.8ex] 
 
\end{tabular} 
\begin{tablenotes}
 \footnotesize
 \justifying \item {\it Notes:}
 This table reports coefficients of the effect of cognitive diversity and average exploratory profile on combinatorial novelty using PKG. Standard errors are cluster robust at the journal level: ***, ** and * indicate significance at the 1\%, 5\% and 10\% levels, respectively. The effects are estimated with an OLS. Variables are field-weighted and constant term, scientific field (Scimago Journal Category), and time-fixed effects are incorporated in all model specifications.
 \end{tablenotes}
 \end{threeparttable}
 }
\label{mesh_cog_nov_fw_pkg}
\end{table} 

\clearpage
\subsection*{Share of Highly Exploratory Profile}

\begin{table}[h!]\footnotesize \centering 
\renewcommand{\arraystretch}{1}
\setlength{\tabcolsep}{0.5pt}
  \caption{Combinatorial Novelty: Cognitive diversity, highly exploratory and exploitative profile (Field-Weighted/ Meshterms)} 
  \label{mesh_share_nov_fw_pkg}
    \scalebox{1}{
	\begin{threeparttable} 
\begin{tabular}{@{\extracolsep{15pt}}lcccc} 
\\[-1.8ex]\hline 
\hline \\[-1.8ex] 
 & \multicolumn{4}{c}{\textit{Dependent variable:}} \\ 
\cline{2-5} 
\\[-1.8ex] & Uzzi & Lee & Foster & Wang  \\ 
\\[-1.8ex] & \multicolumn{1}{c}{(1)} & \multicolumn{1}{c}{(2)} & \multicolumn{1}{c}{(3)} & \multicolumn{1}{c}{(4)}\\ 
\hline \\[-1.8ex] 
 Author inter $_{abs}$ (FW) & 0.045$^{***}$ & 0.115$^{***}$ & 0.007 & 0.002 \\ 
  & (0.011) & (0.010) & (0.013) & (0.008) \\ 
  & & & & \\ 
  Author inter $_{abs}\hat{\mkern6mu}$2  (FW) & 0.013 & -0.032$^{***}$ & 0.030$^{**}$ & 0.027$^{***}$ \\ 
  & (0.010) & (0.010) & (0.012) & (0.007) \\ 
  & & & & \\ 
 Share exploratory & -0.107$^{***}$ & -0.113$^{***}$ & -0.150$^{***}$ & -0.063$^{***}$ \\ 
  & (0.006) & (0.006) & (0.007) & (0.005) \\ 
  & & & & \\ 
 
 Share exploitative & 0.068$^{***}$ & 0.045$^{***}$ & 0.056$^{***}$ & 0.026$^{***}$ \\ 
  & (0.003) & (0.003) & (0.004) & (0.002) \\ 
  & & & & \\ 
 Share exploratory * Share exploitative & 0.185$^{***}$ & 0.189$^{***}$ & 0.254$^{***}$ & 0.106$^{***}$ \\ 
  & (0.016) & (0.013) & (0.016) & (0.012) \\ 
  & & & & \\ 
 \# References & 0.001$^{***}$ & 0.001$^{***}$ & 0.001$^{***}$ & 0.0002$^{***}$ \\ 
  & (0.00005) & (0.00005) & (0.0001) & (0.00003) \\ 
  & & & & \\ 
 \# Meshterms & 0.007$^{***}$ & 0.014$^{***}$ & 0.0004 & 0.029$^{***}$ \\ 
  & (0.001) & (0.001) & (0.0004) & (0.0004) \\ 
  & & & & \\ 
 \# Authors & 0.003$^{***}$ & 0.008$^{***}$ & 0.005$^{***}$ & 0.001$^{***}$ \\ 
  & (0.0004) & (0.0004) & (0.0004) & (0.0003) \\ 
  & & & & \\ 
 SJR & -0.004$^{***}$ & -0.006$^{***}$ & -0.004$^{***}$ & 0.003$^{***}$ \\ 
  & (0.001) & (0.001) & (0.001) & (0.001) \\ 
  & & & & \\ 
  Year & Yes & Yes & Yes & Yes  \\ 
  & & & &  \\ 
  Journal Cat. & Yes & Yes & Yes & Yes \\ 
  & & & &  \\ 
\hline \\[-1.8ex] 
Observations & \multicolumn{1}{c}{661,821} & \multicolumn{1}{c}{1,823,859} & \multicolumn{1}{c}{1,823,859} & \multicolumn{1}{c}{1,823,859} \\ 
R$^{2}$ & \multicolumn{1}{c}{0.033} & \multicolumn{1}{c}{0.086} & \multicolumn{1}{c}{0.019} & \multicolumn{1}{c}{0.152} \\ 
Adjusted R$^{2}$ & \multicolumn{1}{c}{0.032} & \multicolumn{1}{c}{0.086} & \multicolumn{1}{c}{0.019} & \multicolumn{1}{c}{0.152} \\ 
Residual Std. Error & \multicolumn{1}{c}{0.284 } & \multicolumn{1}{c}{0.275 } & \multicolumn{1}{c}{0.299 } & \multicolumn{1}{c}{0.360 } \\ 
F Statistic & \multicolumn{1}{c}{97.014$^{***}$ } & \multicolumn{1}{c}{721.442$^{***}$ } & \multicolumn{1}{c}{149.772$^{***}$ } & \multicolumn{1}{c}{1,383.049$^{***}$ } \\ 
\hline 
\hline \\[-1.8ex] 
 
\end{tabular} 

\begin{tablenotes}
 \footnotesize
 \justifying \item {\it Notes:}
 This table reports coefficients of the effect of cognitive diversity and highly exploratory and exploitative profiles on combinatorial novelty using PKG. Standard errors are cluster robust at the journal level: ***, ** and * indicate significance at the 1\%, 5\% and 10\% levels, respectively. The effects are estimated with an OLS. Variables are field-weighted and constant term, scientific field (Scimago Journal Category), and time-fixed effects are incorporated in all model specifications.
 \end{tablenotes}
 \end{threeparttable}
 }
\end{table} 

\clearpage
\subsection*{Turning points}

\begin{table}[h] \centering 
\renewcommand{\arraystretch}{1.5}
\setlength{\tabcolsep}{0.15pt}
\caption{Turning Points for Combinatorial Novelty  and Scientific Impact}
\label{turning_points_imp}
    \scalebox{0.91}{
	\begin{threeparttable}
\begin{tabular}{@{\extracolsep{35pt}}l cc} 

\midrule
\midrule
Regression & Author intra ${abs}$ (FW) & Author inter ${abs}$ (FW) \\ \hline
Uzzi & 0.318 & 2.725 \\
Lee & 0.229 & 2.441 \\
Foster & 0.244 & 2.521 \\
Wang & 0.038 & 1.75 \\
Shibayama & 2 & 1.203\\ 
\hline
\# Cit. & 0.486 & 0.43 \\
DI1 & 0.75 & 0.875 \\
DI5 & -1.44 & -3.4 \\
DI1nok & 0.166 & -11.75 \\
DeIn & 0.15 & -4.187\\
Breadth & -0.33 & 0.33\\
Depth & -0.052 & 0.083\\
\midrule
\bottomrule
\end{tabular} 

\begin{tablenotes}
 \footnotesize
 \justifying \item {\it Notes:}
 This table reports the turning points of the effect of cognitive diversity and average exploratory profiles on combinatorial novelty and scientific recognition in Table \ref{ref_cog_nov_fw} and \ref{cog_imp_fw_pkg}.
 \end{tablenotes}
 \end{threeparttable}
 }
\end{table} 

\newpage
\subsection*{Regression without field-year weighting}
\begin{table}[h!]\footnotesize \centering
\renewcommand{\arraystretch}{1}
\setlength{\tabcolsep}{0.5pt}
  \caption{Combinatorial Novelty: cognitive diversity and average exploratory profile (References)} 
  \label{ref_cog_nov_pkg} 
    \scalebox{1}{
	\begin{threeparttable}
\begin{tabular}{@{\extracolsep{8pt}}lccccc} 
\\[-1.8ex]\hline 
\hline \\[-1.8ex] 
 & \multicolumn{5}{c}{\textit{Dependent variable:}} \\ 
\cline{2-6}
\\[-1.8ex] &  Uzzi & Lee & Foster & Wang & Shibayama\\ 
\\[-1.8ex] & \multicolumn{1}{c}{(1)} & \multicolumn{1}{c}{(2)} & \multicolumn{1}{c}{(3)} & \multicolumn{1}{c}{(4)} & \multicolumn{1}{c}{(5)}\\ 
\hline \\[-1.8ex] 
Author inter $_{abs}$ & 183.520$^{***}$ & 4.377$^{***}$ & 0.940$^{***}$ & 3.061$^{***}$ & 0.268$^{***}$ \\ 
  & (23.173) & (0.204) & (0.033) & (0.252) & (0.008) \\ 
  & & & & & \\ 
   Author inter $_{abs}\hat{\mkern6mu}$2  & -176.966$^{***}$ & -3.915$^{***}$ & -1.005$^{***}$ & -1.936$^{***}$ & -0.195$^{***}$ \\ 
  & (31.732) & (0.270) & (0.043) & (0.335) & (0.012) \\ 
  & & & & & \\ 
 Author intra $_{abs}$ & 198.281$^{***}$ & 3.825$^{***}$ & 1.052$^{***}$ & 0.095 & 0.226$^{***}$ \\ 
  & (22.235) & (0.222) & (0.074) & (0.365) & (0.011) \\ 
  & & & & & \\ 

 Author intra $_{abs}\hat{\mkern6mu}$2 & -403.151$^{***}$ & -8.090$^{***}$ & -2.057$^{***}$ & 0.130 & -0.153$^{***}$ \\ 
  & (38.759) & (0.381) & (0.107) & (0.619) & (0.018) \\ 
  & & & & & \\ 
 \# References & 0.518$^{***}$ & 0.009$^{***}$ & 0.001$^{***}$ & 0.076$^{***}$ & 0.0004$^{***}$ \\ 
  & (0.072) & (0.0004) & (0.00004) & (0.007) & (0.00002) \\ 
  & & & & & \\ 
 \# Meshterms & 1.287$^{***}$ & 0.025$^{***}$ & 0.003$^{***}$ & -0.043$^{***}$ & 0.001$^{***}$ \\ 
  & (0.119) & (0.002) & (0.0003) & (0.003) & (0.0001) \\ 
  & & & & & \\ 
 \# Authors & 1.371$^{***}$ & 0.025$^{***}$ & 0.004$^{***}$ & 0.005 & 0.002$^{***}$ \\ 
  & (0.113) & (0.001) & (0.0004) & (0.004) & (0.0001) \\ 
  & & & & & \\ 
 SJR & -1.151$^{***}$ & -0.020$^{***}$ & -0.011$^{***}$ & -0.093$^{***}$ & -0.002$^{***}$ \\ 
  & (0.264) & (0.004) & (0.002) & (0.021) & (0.0003) \\ 
  & & & & & \\  
  Year & Yes & Yes & Yes & Yes & Yes  \\ 
  & & & & & \\ 
  Journal Cat. & Yes & Yes & Yes & Yes & Yes \\ 
  & & & & &  \\
\hline \\[-1.8ex] 
Observations & \multicolumn{1}{c}{1,647,446} & \multicolumn{1}{c}{1,815,631} & \multicolumn{1}{c}{1,815,631} & \multicolumn{1}{c}{1,815,631} & \multicolumn{1}{c}{1,809,185} \\ 
R$^{2}$ & \multicolumn{1}{c}{0.020} & \multicolumn{1}{c}{0.168} & \multicolumn{1}{c}{0.151} & \multicolumn{1}{c}{0.158} & \multicolumn{1}{c}{0.253} \\ 
Adjusted R$^{2}$ & \multicolumn{1}{c}{0.020} & \multicolumn{1}{c}{0.168} & \multicolumn{1}{c}{0.151} & \multicolumn{1}{c}{0.158} & \multicolumn{1}{c}{0.253} \\ 
Residual Std. Error & \multicolumn{1}{c}{192.756 } & \multicolumn{1}{c}{1.258 } & \multicolumn{1}{c}{0.235 } & \multicolumn{1}{c}{4.341 } & \multicolumn{1}{c}{0.066} \\ 
F Statistic & \multicolumn{1}{c}{139.319$^{***}$ } & \multicolumn{1}{c}{1,504.472$^{***}$ } & \multicolumn{1}{c}{1,328.955$^{***}$ } & \multicolumn{1}{c}{1,399.846$^{***}$ } & \multicolumn{1}{c}{2,523.818$^{***}$ } \\ 
\hline 
\hline \\[-1.8ex] 

\end{tabular} 
\begin{tablenotes}
 \footnotesize
 \justifying \item {\it Notes:}
 This table reports coefficients of the effect of cognitive diversity and average exploratory profile on combinatorial novelty using PKG. Standard errors are cluster robust at the journal level: ***, ** and * indicate significance at the 1\%, 5\% and 10\% levels, respectively. The effects are estimated with an OLS. The constant term, scientific field (Scimago Journal Category), and time-fixed effects are incorporated in all model specifications.
 \end{tablenotes}
 \end{threeparttable}
 }
\end{table} 
\begin{table}[h!]\footnotesize \centering 
\renewcommand{\arraystretch}{1}
\setlength{\tabcolsep}{0.5pt}
  \caption{Combinatorial Novelty: cognitive diversity and average exploratory profile (Meshterms)} 
  \label{mesh_cog_nov_pkg} 
    \scalebox{1}{
	\begin{threeparttable}
\begin{tabular}{@{\extracolsep{20pt}}lcccc} 
\\[-1.8ex]\hline 
\hline \\[-1.8ex] 
 & \multicolumn{4}{c}{\textit{Dependent variable:}} \\ 
\cline{2-5} 
\\[-1.8ex] & Uzzi & Lee & Foster & Wang \\ 
\\[-1.8ex] & \multicolumn{1}{c}{(1)} & \multicolumn{1}{c}{(2)} & \multicolumn{1}{c}{(3)} & \multicolumn{1}{c}{(4)}\\ 
\hline \\[-1.8ex] 
 Author inter $_{abs}$ & 14.010$^{***}$ & 1.829$^{***}$ & 0.399$^{***}$ & 0.951$^{***}$ \\ 
  & (1.109) & (0.067) & (0.023) & (0.070) \\ 
  & & & & \\ 
   Author inter $_{abs}\hat{\mkern6mu}$2  & -16.049$^{***}$ & -2.002$^{***}$ & -0.495$^{***}$ & -0.915$^{***}$ \\ 
  & (1.477) & (0.087) & (0.034) & (0.086) \\ 
  & & & & \\ 
 Author intra $_{abs}$ & 11.644$^{***}$ & 1.408$^{***}$ & 0.405$^{***}$ & -0.177$^{*}$ \\ 
  & (1.403) & (0.095) & (0.041) & (0.105) \\ 
  & & & & \\ 

 Author intra $_{abs}\hat{\mkern6mu}$2 & -28.595$^{***}$ & -2.603$^{***}$ & -1.066$^{***}$ & -0.578$^{***}$ \\ 
  & (2.138) & (0.140) & (0.063) & (0.154) \\ 
  & & & & \\ 
 \# References & 0.038$^{***}$ & 0.002$^{***}$ & 0.001$^{***}$ & 0.001$^{***}$ \\ 
  & (0.002) & (0.0001) & (0.00004) & (0.0001) \\ 
  & & & & \\ 
 \# Meshterms & -0.022$^{*}$ & 0.028$^{***}$ & 0.001$^{***}$ & 0.058$^{***}$ \\ 
  & (0.013) & (0.001) & (0.0003) & (0.001) \\ 
  & & & & \\ 
 \# Authors & 0.012 & 0.011$^{***}$ & 0.002$^{***}$ & -0.0001 \\ 
  & (0.008) & (0.001) & (0.0003) & (0.001) \\ 
  & & & & \\ 
 SJR & -0.052 & -0.011$^{***}$ & -0.002$^{***}$ & 0.015$^{***}$ \\ 
  & (0.032) & (0.002) & (0.001) & (0.003) \\ 
  & & & & \\ 
  Year & Yes & Yes & Yes & Yes  \\ 
  & & & &  \\ 
  Journal Cat. & Yes & Yes & Yes & Yes \\ 
  & & & &  \\ 
\hline \\[-1.8ex] 
Observations & \multicolumn{1}{c}{661,832} & \multicolumn{1}{c}{1,823,889} & \multicolumn{1}{c}{1,823,889} & \multicolumn{1}{c}{1,823,889} \\ 
R$^{2}$ & \multicolumn{1}{c}{0.064} & \multicolumn{1}{c}{0.179} & \multicolumn{1}{c}{0.120} & \multicolumn{1}{c}{0.174} \\ 
Adjusted R$^{2}$ & \multicolumn{1}{c}{0.063} & \multicolumn{1}{c}{0.179} & \multicolumn{1}{c}{0.120} & \multicolumn{1}{c}{0.174} \\ 
Residual Std. Error & \multicolumn{1}{c}{7.929 } & \multicolumn{1}{c}{0.536 } & \multicolumn{1}{c}{0.206 } & \multicolumn{1}{c}{0.716 } \\ 
F Statistic & \multicolumn{1}{c}{193.801$^{***}$ } & \multicolumn{1}{c}{1,638.907$^{***}$ } & \multicolumn{1}{c}{1,020.027$^{***}$ } & \multicolumn{1}{c}{1,586.294$^{***}$ } \\ 
\hline 
\hline \\[-1.8ex] 
 
\end{tabular} 
\begin{tablenotes}
 \footnotesize
 \justifying \item {\it Notes:}
 This table reports coefficients of the effect of cognitive diversity and average exploratory profile on combinatorial novelty using PKG. Standard errors are cluster robust at the journal level: ***, ** and * indicate significance at the 1\%, 5\% and 10\% levels, respectively. The effects are estimated with an OLS. The constant term, scientific field (Scimago Journal Category), and time-fixed effects are incorporated in all model specifications.
 \end{tablenotes}
 \end{threeparttable}
 }
\end{table} 
\begin{table}[h!]\footnotesize \centering 
\renewcommand{\arraystretch}{1}
\setlength{\tabcolsep}{0.5pt}
  \caption{Scientific recognition: cognitive diversity and average exploratory profile} 
  \label{cog_imp_pkg} 
    \scalebox{0.9}{
	\begin{threeparttable}
\begin{tabular}{@{\extracolsep{5pt}}lccccccc} 
\\[-1.8ex]\hline 
\hline \\[-1.8ex] 
 & \multicolumn{7}{c}{\textit{Dependent variable:}} \\ 
\cline{2-8} 
\\[-1.8ex] & \# cit. & DI1 & DI5 & DI1nok & DeIn & Breadth & Depth \\ 
\\[-1.8ex] & \multicolumn{1}{c}{(1)} & \multicolumn{1}{c}{(2)} & \multicolumn{1}{c}{(3)} & \multicolumn{1}{c}{(4)} & \multicolumn{1}{c}{(5)} & \multicolumn{1}{c}{(6)} & \multicolumn{1}{c}{(7)}\\ 
\hline \\[-1.8ex] 
 Author inter $_{abs}$ & 39.622$^{***}$ & -0.019$^{***}$ & -0.043$^{***}$ & 0.640$^{***}$ & -3.961$^{***}$ & 0.044$^{*}$ & -0.080$^{***}$ \\ 
  & (12.175) & (0.006) & (0.008) & (0.050) & (0.306) & (0.023) & (0.024) \\ 
  & & & & & & & \\ 
  Author inter $_{abs}\hat{\mkern6mu}$2  & -55.784$^{***}$ & 0.034$^{***}$ & 0.068$^{***}$ & -0.485$^{***}$ & 3.848$^{***}$ & -0.023 & 0.057$^{*}$ \\ 
  & (15.267) & (0.008) & (0.010) & (0.066) & (0.379) & (0.032) & (0.033) \\ 
  & & & & & & & \\ 
 Author intra $_{abs}$ & 99.833$^{***}$ & -0.064$^{***}$ & -0.067$^{***}$ & -0.090 & -2.059$^{***}$ & 0.111$^{***}$ & -0.033 \\ 
  & (12.635) & (0.007) & (0.008) & (0.073) & (0.385) & (0.032) & (0.033) \\ 
  & & & & & & & \\ 
 
 Author intra $_{abs}\hat{\mkern6mu}$2 & -130.168$^{***}$ & 0.069$^{***}$ & 0.094$^{***}$ & 0.138 & 2.499$^{***}$ & -0.021 & -0.141$^{***}$ \\ 
  & (16.543) & (0.010) & (0.012) & (0.105) & (0.541) & (0.047) & (0.049) \\ 
  & & & & & & & \\ 
 \# References & 0.681$^{***}$ & -0.0002$^{***}$ & -0.0004$^{***}$ & -0.003$^{***}$ & 0.023$^{***}$ & -0.0002$^{***}$ & 0.001$^{***}$ \\ 
  & (0.027) & (0.00001) & (0.00002) & (0.0001) & (0.001) & (0.00004) & (0.0001) \\ 
  & & & & & & & \\ 
 \# Meshterms & 0.338$^{***}$ & -0.0004$^{***}$ & -0.001$^{***}$ & -0.006$^{***}$ & 0.019$^{***}$ & -0.003$^{***}$ & 0.005$^{***}$ \\ 
  & (0.080) & (0.00004) & (0.0001) & (0.0005) & (0.002) & (0.0002) & (0.0003) \\ 
  & & & & & & & \\ 
 \# Authors & 3.405$^{***}$ & -0.0004$^{***}$ & -0.0002$^{***}$ & -0.009$^{***}$ & 0.023$^{***}$ & -0.008$^{***}$ & 0.010$^{***}$ \\ 
  & (0.365) & (0.00003) & (0.00005) & (0.0005) & (0.002) & (0.0003) & (0.0004) \\ 
  & & & & & & & \\ 
 SJR & 16.482$^{***}$ & -0.001$^{***}$ & 0.001$^{***}$ & -0.013$^{***}$ & 0.025$^{***}$ & -0.023$^{***}$ & 0.025$^{***}$ \\ 
  & (1.269) & (0.0001) & (0.0002) & (0.002) & (0.009) & (0.003) & (0.003) \\ 
  & & & & & & & \\ 
  Year & Yes & Yes & Yes & Yes & Yes &  Yes  \\ 
  & & & & & & & \\ 
  Journal Cat. & Yes & Yes & Yes & Yes & Yes &  Yes  \\ 
  & & & & & & & \\ 
\hline \\[-1.8ex] 
Observations & \multicolumn{1}{c}{1,826,237} & \multicolumn{1}{c}{1,826,237} & \multicolumn{1}{c}{1,826,237} & \multicolumn{1}{c}{1,826,237} & \multicolumn{1}{c}{1,826,237} & \multicolumn{1}{c}{1,826,237} & \multicolumn{1}{c}{1,826,237} \\ 
R$^{2}$ & \multicolumn{1}{c}{0.116} & \multicolumn{1}{c}{0.042} & \multicolumn{1}{c}{0.077} & \multicolumn{1}{c}{0.133} & \multicolumn{1}{c}{0.238} & \multicolumn{1}{c}{0.107} & \multicolumn{1}{c}{0.159} \\ 
Adjusted R$^{2}$ & \multicolumn{1}{c}{0.116} & \multicolumn{1}{c}{0.042} & \multicolumn{1}{c}{0.077} & \multicolumn{1}{c}{0.133} & \multicolumn{1}{c}{0.238} & \multicolumn{1}{c}{0.107} & \multicolumn{1}{c}{0.158} \\ 
Residual Std. Error  & \multicolumn{1}{c}{126.203} & \multicolumn{1}{c}{0.056} & \multicolumn{1}{c}{0.061} & \multicolumn{1}{c}{0.467} & \multicolumn{1}{c}{1.591} & \multicolumn{1}{c}{0.250} & \multicolumn{1}{c}{0.241} \\ 
F Statistic  & \multicolumn{1}{c}{984.300$^{***}$} & \multicolumn{1}{c}{328.932$^{***}$} & \multicolumn{1}{c}{626.917$^{***}$} & \multicolumn{1}{c}{1,151.082$^{***}$} & \multicolumn{1}{c}{2,343.227$^{***}$} & \multicolumn{1}{c}{904.045$^{***}$} & \multicolumn{1}{c}{1,416.198$^{***}$} \\ 
\hline 
\hline \\[-1.8ex] 
 
\end{tabular} 
\begin{tablenotes}
 \footnotesize
 \justifying \item {\it Notes:}
 This table reports coefficients of the effect of cognitive diversity and average exploratory profile on scientific recognition using PKG. Standard errors are cluster robust at the journal level: ***, ** and * indicate significance at the 1\%, 5\% and 10\% levels, respectively. The effects are estimated with an OLS. The constant term, scientific field (Scimago Journal Category), and time-fixed effects are incorporated in all model specifications.
 \end{tablenotes}
 \end{threeparttable}
 }
\end{table}

\begin{table}[h!]\footnotesize \centering 
\renewcommand{\arraystretch}{1}
\setlength{\tabcolsep}{0.5pt}
  \caption{Combinatorial Novelty: Cognitive diversity, highly exploratory and exploitative profile (References)} 
  \label{ref_share_nov_pkg} 
    \scalebox{0.9}{
	\begin{threeparttable}
\begin{tabular}{@{\extracolsep{5pt}}lccccc} 
\\[-1.8ex]\hline 
\hline \\[-1.8ex] 
 & \multicolumn{5}{c}{\textit{Dependent variable:}} \\ 
\cline{2-6} 
\\[-1.8ex] & Uzzi & Lee & Foster & Wang & Shibayama \\ 
\\[-1.8ex] & \multicolumn{1}{c}{(1)} & \multicolumn{1}{c}{(2)} & \multicolumn{1}{c}{(3)} & \multicolumn{1}{c}{(4)} & \multicolumn{1}{c}{(5)}\\ 
\hline \\[-1.8ex] 
  Author inter $_{abs}$ & 280.445$^{***}$ & 6.383$^{***}$ & 1.533$^{***}$ & 1.696$^{***}$ & 0.430$^{***}$ \\ 
  & (30.366) & (0.225) & (0.066) & (0.347) & (0.011) \\ 
  & & & & & \\ 
  Author inter $_{abs}\hat{\mkern6mu}$2  & -311.743$^{***}$ & -6.771$^{***}$ & -1.834$^{***}$ & -0.650 & -0.350$^{***}$ \\ 
  & (39.387) & (0.298) & (0.082) & (0.493) & (0.015) \\ 
  & & & & & \\ 
 Share exploratory & -22.978$^{***}$ & -0.416$^{***}$ & -0.087$^{***}$ & -0.328$^{***}$ & 0.009$^{***}$ \\ 
  & (3.028) & (0.029) & (0.005) & (0.034) & (0.001) \\ 
  & & & & & \\ 

 Share exploitative & 8.714$^{***}$ & 0.234$^{***}$ & 0.048$^{***}$ & -0.468$^{***}$ & -0.021$^{***}$ \\ 
  & (1.809) & (0.015) & (0.004) & (0.082) & (0.001) \\ 
  & & & & & \\ 
 Share exploratory * Share exploitative & 29.023$^{***}$ & 0.541$^{***}$ & 0.186$^{***}$ & 0.208 & -0.052$^{***}$ \\ 
  & (8.047) & (0.084) & (0.013) & (0.129) & (0.004) \\ 
  & & & & & \\ 
 \# References & 0.514$^{***}$ & 0.009$^{***}$ & 0.001$^{***}$ & 0.076$^{***}$ & 0.0004$^{***}$ \\ 
  & (0.073) & (0.0004) & (0.00004) & (0.007) & (0.00002) \\ 
  & & & & & \\ 
 \# Meshterms & 1.292$^{***}$ & 0.025$^{***}$ & 0.004$^{***}$ & -0.042$^{***}$ & 0.001$^{***}$ \\ 
  & (0.119) & (0.002) & (0.0003) & (0.003) & (0.0001) \\ 
  & & & & & \\ 
 \# Authors & 1.472$^{***}$ & 0.027$^{***}$ & 0.005$^{***}$ & 0.002 & 0.001$^{***}$ \\ 
  & (0.118) & (0.001) & (0.0004) & (0.003) & (0.0001) \\ 
  & & & & & \\ 
 SJR & -1.206$^{***}$ & -0.022$^{***}$ & -0.011$^{***}$ & -0.090$^{***}$ & -0.002$^{***}$ \\ 
  & (0.260) & (0.004) & (0.002) & (0.020) & (0.0003) \\ 
  & & & & & \\ 
  Year & Yes & Yes & Yes & Yes & Yes  \\ 
  & & & & & \\ 
  Journal Cat. & Yes & Yes & Yes & Yes & Yes \\ 
  & & & & &  \\
\hline \\[-1.8ex] 
Observations & \multicolumn{1}{c}{1,647,446} & \multicolumn{1}{c}{1,815,631} & \multicolumn{1}{c}{1,815,631} & \multicolumn{1}{c}{1,815,631} & \multicolumn{1}{c}{1,809,185} \\ 
R$^{2}$ & \multicolumn{1}{c}{0.020} & \multicolumn{1}{c}{0.167} & \multicolumn{1}{c}{0.150} & \multicolumn{1}{c}{0.158} & \multicolumn{1}{c}{0.252} \\ 
Adjusted R$^{2}$ & \multicolumn{1}{c}{0.020} & \multicolumn{1}{c}{0.167} & \multicolumn{1}{c}{0.150} & \multicolumn{1}{c}{0.158} & \multicolumn{1}{c}{0.252} \\ 
Residual Std. Error & \multicolumn{1}{c}{192.763 } & \multicolumn{1}{c}{1.258 } & \multicolumn{1}{c}{0.235 } & \multicolumn{1}{c}{4.340 } & \multicolumn{1}{c}{0.067 } \\ 
F Statistic & \multicolumn{1}{c}{138.223$^{***}$ } & \multicolumn{1}{c}{1,493.115$^{***}$ } & \multicolumn{1}{c}{1,310.908$^{***}$ } & \multicolumn{1}{c}{1,399.600$^{***}$ } & \multicolumn{1}{c}{2,503.790$^{***}$} \\ 
\hline 
\hline \\[-1.8ex] 

\end{tabular} 

\begin{tablenotes}
 \footnotesize
 \justifying \item {\it Notes:}
 This table reports coefficients of the effect of cognitive diversity and highly exploratory and exploitative profiles on combinatorial novelty using PKG. Standard errors are cluster robust at the journal level: ***, ** and * indicate significance at the 1\%, 5\% and 10\% levels, respectively. The effects are estimated with an OLS. The constant term, scientific field (Scimago Journal Category), and time-fixed effects are incorporated in all model specifications.
 \end{tablenotes}
 \end{threeparttable}
 }
\end{table} 
\begin{table}[h!]\footnotesize \centering 
\renewcommand{\arraystretch}{1}
\setlength{\tabcolsep}{0.5pt}
  \caption{Combinatorial Novelty: Cognitive diversity, highly exploratory and exploitative profile (Meshterms)} 
  \label{mesh_share_nov_pkg} 
    \scalebox{1}{
	\begin{threeparttable}
\begin{tabular}{@{\extracolsep{8pt}}lcccc} 
\\[-1.8ex]\hline 
\hline \\[-1.8ex] 
 & \multicolumn{4}{c}{\textit{Dependent variable:}} \\ 
\cline{2-5} 
\\[-1.8ex] & Uzzi & Lee & Foster & Wang \\ 
\\[-1.8ex] & \multicolumn{1}{c}{(1)} & \multicolumn{1}{c}{(2)} & \multicolumn{1}{c}{(3)} & \multicolumn{1}{c}{(4)}\\ 
\hline \\[-1.8ex] 
Author inter $_{abs}$ & 23.097$^{***}$ & 2.721$^{***}$ & 0.573$^{***}$ & 0.687$^{***}$ \\ 
  & (1.222) & (0.095) & (0.036) & (0.103) \\ 
  & & & & \\ 
   Author inter $_{abs}\hat{\mkern6mu}$2  & -28.625$^{***}$ & -3.128$^{***}$ & -0.798$^{***}$ & -0.852$^{***}$ \\ 
  & (1.611) & (0.118) & (0.047) & (0.125) \\ 
  & & & & \\ 
 Share exploratory & -0.852$^{***}$ & -0.091$^{***}$ & -0.068$^{***}$ & -0.073$^{***}$ \\ 
  & (0.100) & (0.007) & (0.004) & (0.007) \\ 
  & & & & \\ 

 Share exploitative & 1.977$^{***}$ & 0.078$^{***}$ & 0.045$^{***}$ & 0.057$^{***}$ \\ 
  & (0.092) & (0.007) & (0.003) & (0.006) \\ 
  & & & & \\ 
 Share exploratory * Share exploitative & 1.251$^{***}$ & 0.083$^{***}$ & 0.130$^{***}$ & 0.135$^{***}$ \\ 
  & (0.393) & (0.023) & (0.010) & (0.025) \\ 
  & & & & \\ 
 \# References & 0.038$^{***}$ & 0.002$^{***}$ & 0.001$^{***}$ & 0.001$^{***}$ \\ 
  & (0.001) & (0.0001) & (0.00004) & (0.0001) \\ 
  & & & & \\ 
 \# Meshterms & -0.022$^{*}$ & 0.028$^{***}$ & 0.001$^{***}$ & 0.058$^{***}$ \\ 
  & (0.013) & (0.001) & (0.0003) & (0.001) \\ 
  & & & & \\ 
 \# Authors & 0.028$^{***}$ & 0.012$^{***}$ & 0.003$^{***}$ & 0.001 \\ 
  & (0.008) & (0.001) & (0.0003) & (0.001) \\ 
  & & & & \\ 
 SJR & -0.064$^{**}$ & -0.012$^{***}$ & -0.003$^{***}$ & 0.015$^{***}$ \\ 
  & (0.031) & (0.002) & (0.001) & (0.003) \\ 
  & & & & \\ 
  Year & Yes & Yes & Yes & Yes  \\ 
  & & & &  \\ 
  Journal Cat. & Yes & Yes & Yes & Yes \\ 
  & & & &  \\ 
\hline \\[-1.8ex] 
Observations & \multicolumn{1}{c}{661,832} & \multicolumn{1}{c}{1,823,889} & \multicolumn{1}{c}{1,823,889} & \multicolumn{1}{c}{1,823,889} \\ 
R$^{2}$ & \multicolumn{1}{c}{0.065} & \multicolumn{1}{c}{0.179} & \multicolumn{1}{c}{0.119} & \multicolumn{1}{c}{0.174} \\ 
Adjusted R$^{2}$ & \multicolumn{1}{c}{0.065} & \multicolumn{1}{c}{0.179} & \multicolumn{1}{c}{0.119} & \multicolumn{1}{c}{0.174} \\ 
Residual Std. Error & \multicolumn{1}{c}{7.923 } & \multicolumn{1}{c}{0.537 } & \multicolumn{1}{c}{0.206 } & \multicolumn{1}{c}{0.716 } \\ 
F Statistic & \multicolumn{1}{c}{197.576$^{***}$ } & \multicolumn{1}{c}{1,631.871$^{***}$ } & \multicolumn{1}{c}{1,014.063$^{***}$ } & \multicolumn{1}{c}{1,574.819$^{***}$ } \\ 
\hline 
\hline \\[-1.8ex] 
 
\end{tabular} 

\begin{tablenotes}
 \footnotesize
 \justifying \item {\it Notes:}
 This table reports coefficients of the effect of cognitive diversity and highly exploratory and exploitative profiles on combinatorial novelty using PKG. Standard errors are cluster robust at the journal level: ***, ** and * indicate significance at the 1\%, 5\% and 10\% levels, respectively. The effects are estimated with an OLS. The constant term, scientific field (Scimago Journal Category), and time-fixed effects are incorporated in all model specifications.
 \end{tablenotes}
 \end{threeparttable}
 }
\end{table} 
\begin{table}[h!]\footnotesize \centering 
\renewcommand{\arraystretch}{1}
\setlength{\tabcolsep}{0.5pt}
  \caption{Scientific recognition: cognitive diversity, highly exploratory and exploitative profile} 
  \label{share_imp_pkg} 
    \scalebox{0.79}{
	\begin{threeparttable}
\begin{tabular}{@{\extracolsep{3pt}}lccccccc} 
\\[-1.8ex]\hline 
\hline \\[-1.8ex] 
 & \multicolumn{7}{c}{\textit{Dependent variable:}} \\ 
\cline{2-8} 
\\[-1.8ex] &  \# cit. & DI1 & DI5 & DI1nok & DeIn & Breadth & Depth \\ 
\\[-1.8ex] & \multicolumn{1}{c}{(1)} & \multicolumn{1}{c}{(2)} & \multicolumn{1}{c}{(3)} & \multicolumn{1}{c}{(4)} & \multicolumn{1}{c}{(5)} & \multicolumn{1}{c}{(6)} & \multicolumn{1}{c}{(7)}\\ 
\hline \\[-1.8ex] 
 Author inter $_{abs}$ & 114.316$^{***}$ & -0.088$^{***}$ & -0.115$^{***}$ & 0.313$^{***}$ & -4.628$^{***}$ & 0.114$^{***}$ & -0.095$^{***}$ \\ 
  & (13.961) & (0.009) & (0.011) & (0.084) & (0.556) & (0.030) & (0.033) \\ 
  & & & & & & & \\ 
   Author inter $_{abs}$$\hat{\mkern6mu}$2 & -122.966$^{***}$ & 0.101$^{***}$ & 0.143$^{***}$ & -0.193$^{*}$ & 4.793$^{***}$ & -0.093$^{**}$ & 0.041 \\ 
  & (16.903) & (0.011) & (0.013) & (0.103) & (0.669) & (0.039) & (0.043) \\ 
  & & & & & & & \\ 
 Share exploratory & -2.597$^{**}$ & -0.006$^{***}$ & -0.005$^{***}$ & -0.057$^{***}$ & 0.176$^{***}$ & 0.007$^{**}$ & -0.009$^{***}$ \\ 
  & (1.100) & (0.001) & (0.001) & (0.005) & (0.018) & (0.003) & (0.003) \\ 
  & & & & & & & \\ 

 Share exploitative & 3.759$^{***}$ & -0.005$^{***}$ & -0.008$^{***}$ & -0.085$^{***}$ & 0.284$^{***}$ & -0.021$^{***}$ & 0.027$^{***}$ \\ 
  & (1.085) & (0.0005) & (0.001) & (0.004) & (0.015) & (0.002) & (0.002) \\ 
  & & & & & & & \\ 
 Share exploratory * Share exploitative & -23.770$^{***}$ & 0.019$^{***}$ & 0.014$^{***}$ & 0.115$^{***}$ & -0.181$^{***}$ & 0.031$^{***}$ & -0.017$^{*}$ \\ 
  & (3.529) & (0.002) & (0.002) & (0.017) & (0.064) & (0.010) & (0.010) \\ 
  & & & & & & & \\ 
 \# References & 0.679$^{***}$ & -0.0002$^{***}$ & -0.0004$^{***}$ & -0.003$^{***}$ & 0.023$^{***}$ & -0.0002$^{***}$ & 0.001$^{***}$ \\ 
  & (0.027) & (0.00001) & (0.00002) & (0.0001) & (0.001) & (0.00004) & (0.0001) \\ 
  & & & & & & & \\ 
 \# Meshterms & 0.337$^{***}$ & -0.0004$^{***}$ & -0.001$^{***}$ & -0.006$^{***}$ & 0.019$^{***}$ & -0.003$^{***}$ & 0.005$^{***}$ \\ 
  & (0.080) & (0.00004) & (0.0001) & (0.0005) & (0.002) & (0.0002) & (0.0003) \\ 
  & & & & & & & \\ 
 \# Authors & 3.430$^{***}$ & -0.0004$^{***}$ & -0.0003$^{***}$ & -0.010$^{***}$ & 0.025$^{***}$ & -0.009$^{***}$ & 0.011$^{***}$ \\ 
  & (0.364) & (0.00003) & (0.00005) & (0.0005) & (0.002) & (0.0003) & (0.0004) \\ 
  & & & & & & & \\ 
 SJR & 16.427$^{***}$ & -0.001$^{***}$ & 0.001$^{***}$ & -0.012$^{***}$ & 0.024$^{**}$ & -0.023$^{***}$ & 0.025$^{***}$ \\ 
  & (1.268) & (0.0001) & (0.0002) & (0.002) & (0.009) & (0.003) & (0.003) \\ 
  & & & & & & & \\ 
  Year & Yes & Yes & Yes & Yes & Yes &  Yes & Yes \\ 
  & & & & & & & \\ 
  Journal Cat. & Yes & Yes & Yes & Yes & Yes &  Yes & Yes  \\ 
  & & & & & & & \\ 
\hline \\[-1.8ex] 
Observations & \multicolumn{1}{c}{1,826,237} & \multicolumn{1}{c}{1,826,237} & \multicolumn{1}{c}{1,826,237} & \multicolumn{1}{c}{1,826,237} & \multicolumn{1}{c}{1,826,237} & \multicolumn{1}{c}{1,826,237} & \multicolumn{1}{c}{1,826,237} \\ 
R$^{2}$ & \multicolumn{1}{c}{0.116} & \multicolumn{1}{c}{0.042} & \multicolumn{1}{c}{0.078} & \multicolumn{1}{c}{0.134} & \multicolumn{1}{c}{0.239} & \multicolumn{1}{c}{0.107} & \multicolumn{1}{c}{0.159} \\ 
Adjusted R$^{2}$ & \multicolumn{1}{c}{0.116} & \multicolumn{1}{c}{0.042} & \multicolumn{1}{c}{0.078} & \multicolumn{1}{c}{0.134} & \multicolumn{1}{c}{0.239} & \multicolumn{1}{c}{0.107} & \multicolumn{1}{c}{0.159} \\ 
Residual Std. Error  & \multicolumn{1}{c}{126.204} & \multicolumn{1}{c}{0.056} & \multicolumn{1}{c}{0.061} & \multicolumn{1}{c}{0.467} & \multicolumn{1}{c}{1.590} & \multicolumn{1}{c}{0.250} & \multicolumn{1}{c}{0.241} \\ 
F Statistic  & \multicolumn{1}{c}{980.132$^{***}$} & \multicolumn{1}{c}{328.824$^{***}$} & \multicolumn{1}{c}{630.089$^{***}$} & \multicolumn{1}{c}{1,160.693$^{***}$} & \multicolumn{1}{c}{2,346.784$^{***}$} & \multicolumn{1}{c}{901.029$^{***}$} & \multicolumn{1}{c}{1,411.683$^{***}$} \\ 
\hline 
\hline \\[-1.8ex] 

\end{tabular} 
\begin{tablenotes}
 \footnotesize
 \justifying \item {\it Notes:}
 This table reports coefficients of the effect of cognitive diversity and highly exploratory and exploitative profiles on scientific recognition using PKG. Standard errors are cluster robust at the journal level: ***, ** and * indicate significance at the 1\%, 5\% and 10\% levels, respectively. The effects are estimated with an OLS. The constant term, scientific field (Scimago Journal Category), and time-fixed effects are incorporated in all model specifications.
 \end{tablenotes}
 \end{threeparttable}
 }
\end{table}

\end{document}